\documentclass[useAMS,usenatbib]{mn2e}
\usepackage{epsfig}
\title[Bispectrum]{The biphase explained: understanding the
  asymmetries in coupled Fourier components of astronomical
  timeseries}

\author[Maccarone et al.]{Thomas J. Maccarone\\ Department of Physics, Box 41051, Science Building, Texas Tech University, Lubbock TX 79409-1051\\ School of Physics and Astronomy, University of Southampton, SO16 4ES\\email:thomas.maccarone@ttu.edu}

\begin{document}
\def\ltsim{\mathrel{\rlap{\lower 3pt\hbox{$\sim$}}
        \raise 2.0pt\hbox{$<$}}}
\def\gtsim{\mathrel{\rlap{\lower 3pt\hbox{$\sim$}}
        \raise 2.0pt\hbox{$>$}}}

\date{}

\pagerange{\pageref{firstpage}--\pageref{lastpage}} \pubyear{}

\maketitle

\label{firstpage}

\begin{abstract}

We make the first attempt to estimate and interpret the biphase
data for astronomical time series.  The biphase is the phase of the
bispectrum, which is the Fourier domain equivalent of the three-point
correlation function.  The bispectrum measures two key nonlinear
properties of a time series -- its reversability in time, and the
symmetry about the mean of its flux distribution -- for triplets of
frequencies.  Like other Fourier methods, it is especially valuable
for working with time series which contain large numbers of cycles at
the period of interest, but in which the signal-to-noise at a given
frequency is small in any individual cycle, either because of
measurement errors, or because of the contributions from signals at
other frequencies.  This has long been the case for studies of X-ray
binaries, but is increasingly becoming true for stellar variability
(both intrinsic and due to planetary transits) in the Kepler era.  We
attempt in this paper also to present some simple examples to give a
more intuitive understanding of the meaning of the bispectrum to
readers, in order to help to understand where it may be applicable in
astronomy.  In particular, we give illustrative examples of what
biphases may be shown by common astrophysical time series such as
pulsars, eclipsers, stars in the instability strip, and solar flares.
We then discuss applications to the biphase data for understanding the
shapes of the quasi-periodic oscillations of GRS~1915+105 and the
coupling of the quasi-periodic oscillations to the power-law noise in
that system.
\end{abstract}


\section{Introduction}

Astronomy is one of the first sciences to make use of time series analysis,
with the studies of the orbits of planets in the solar system (see e.g. Way et
al. 2012 and references within for some examples).  Perhaps because of the
historical emphasis on orbits, astronomical time series analysis has
traditonally focused on the frequencies at which the strongest variability is
found, with comparitively less emphasis on the phases of different Fourier
components of time series.

In a variety of systems in nature, and in laboratory studies of
dynamical systems, power spectra of sources reveal strong variability
over a wide range of frequencies.  In order to be able to demonstrate
that power exists on a large range of timescales, one ideally will
have time series on which to work which are uninterrupted (or at least
regularly sampled) and which are long relative to the timescales of
interest.  In X-ray binaries, the fast timescales of variation and
strong variability allow one to probe a wide range of timescales
effectively, and it has been known for about four decades that
aperiodic variability can be strong in these objects (e.g. Terrell
1972).  Time series of magnetograms from active regions of the Sun,
which can also be made at high cadence over long time spans, also show
power spectra well modelled by power laws over wide ranges of
frequencies, rather than by power at a few discrete frequencies
(Abramenko 2005).  In recent years, the Kepler satellite has taken
long, nearly uninterrupted time series of many other stars, and
components in the power spectra of solar-type stars which span a broad
range of frequencies have been observed (e.g. Jiang et al. 2011).

Simple tools exist for studying the time profiles of oscillations when
the variability is strictly periodic (albeit, perhaps non-sinusoidal).
In cases of bright sources, with variations much larger than the noise
level on individual data points, one can simply examine the raw time
series in the time domain.  When larger noise components are present
(whether they are physical, such as the noise due to stellar activity
on a star with planetary transits, or are simply noise due to
measurement uncertainties), one can fold the time series on the period
and see the mean profile.  What has generally not been done in
astronomy is to examine the phase couplings of aperiodic variability.

In the cases where phase dependences are studied, often, the emphasis
is on lags between different photon wavelengths (as e.g. in
reverberation mapping of active galactic nuclei -- e.g. Edelson \&
Krolik 1998 or studies of time lags in X-ray binaries -- e.g. Nowak et
al. 1999) -- although, in some cases, some information about the
nonlinearity of a system can be obtained solely through studies of the
power spectrum and the cross-spectrum or cross correlation function
(e.g. Maccarone, Coppi \& Poutanen 2000; Shaposhnikov 2012).  The
reasons for this are twofold.  First, under some circumstances the
meaning of the phase lag between two wavelengths of light can have an
immediately obvious interpretation.  For example, in the case of
reverberation mapping, it gives the added light travel time by taking
a route that passes through the line emission region.

Secondly, making measurements of nonlinear variability in systems
which possibly have red noise contributions requires a large number of
high quality independent measurements of the Fourier spectrum (or some
alternative statistical measure of the Fourier spectrum).  This is
rarely the case in astronomy.  Furthermore, the real advantages of
non-linearity analyses in the Fourier domain are seen when the
signal-to-noise on individual measurements is poor, but a very large
number of measurements exist; and/or there is substantial aperiodic
variability, or there are a very large number of frequencies
contributing to the variability, so that simple folding of the data on
a characteristic period does not capture all that is happening in the
system.

X-ray binaries represent a particularly good example of a class of
systems which are particularly ripe for sophisticated nonlinearity
analyses.  They show variability on a wide range of timescales.  They
are typically the subjects of very low background rate observations
where individual photons are counted and, with many X-ray
observatories, the count rate per time resolution element is
significantly less than unity.  In recent years satellites such as
Kepler and Corot have obtained very long, high precision uninterrupted
observations of bright stars, which are likely to be affected by some
combination of asteroseismic modes, planetary transits, coronally
activity and atmospheric turbulence (e.g Jiang et al. 2011), and which
also allow the study of the evolution of the rapid variability of
cataclysmic variables (Scaringi et al. 2012).

There have been some attempts made to characterize and understand the
nonlinear variability of X-ray binaries.  In the very early era of
X-ray astronomy, some attempts were made to measure, for example, the
asymmetries of light curves (Priedhorsky et al.1979).  More recently,
that there is some nonlinearity was proved by the presence of an
rms-flux relation (Uttley \& McHardy 2001).

In this paper, we present a more detailed treatment than has been
presented in the past of what can be learned from use of the
bispectrum.  Because the topic is fairly new to astronomical time
series analysis, we will take a more pedagogical tone than is taken in
most typical papers in astronomy, and will develop some ideas already
well known in other fields of research.  We will show, in particular,
that the bispectrum presents a good means for determining whether a
time series is reversible (in a statistical sense) and whether a time
series has a symmetric flux distribution.

\section{The bispectrum: a tutorial}

The bispectrum is an example of a higher order time series analysis
technique which can be used to understand the phase correlations in a
single time series.  Successful applications have been made in studies
of brain waves (e.g. Gajraj et al. 1998), of speech patterns (Fackrell
1997), of vibrations of machinery (Rivola \& White 1998), plasma
physics (van Milligen et al. 1995), and, ocean waves, for which it was
first developed (Hasselman et al. 1963).  Spatial, rather than
temporal, bispectra have been studied widely in astronomy, for
purposes of undertanding the non-gaussianity of the cosmic microwave
background (e.g. Kamionkowski et al. 2011).  A large fraction of the
literature on the bispectrum was developed for the study of ocean
waves, and we will draw heavily on what has already been developed in
that field for building up our understanding of what we can learn from
the bispectrum.

The bispectrum is the first in a series of polyspectra -- analogies to
the classical Fourier spectrum which take into account more than one
timescale.  The bispectrum of two frequencies, $k$ and $l$, $B(k,l)$
is defined by:

\begin{equation}
B(k,l)=\frac{1}{K} \sum_{i=0}^{K-1} X_i(k)X_i(l)X^*_i(k+l),
\label{bispeceqn}
\end{equation}

where there are $K$ segments to a time series, and $X_i(f)$ denotes
the Fourier transform of the $i$th segment of the light curve at
frequency $f$.  The asterisk is used in the final term to denote that
a complex conjugate is being taken.

One can see that, in order to produce useful measurements of the
bispectrum, one needs to have a large number of independent
measurements of the time series, each with high signal to noise, and
with the power spectrum stationary over the duration of the
observations.  The expectation value of the bispectrum is unaffected
by Gaussian noise, but its value can be strongly affected by Poisson
noise (e.g. Uttley et al. 2005), since Poisson noise is nonlinear.

The bispectrum is related to two quantities of a time series, the
skewness and the asymmetry.  The skewness is related to the mean cube
of the values of the data points in a distribution in the same way
that the variance is related to the mean square of the data points in
a distribution -- i.e. it is the third moment of the flux
distribution.  The asymmetry is related to the directionality of the
time series, in a manner similar, but not identical, to the time
skewness statistic developed by Priedhorsky et al. (1979) which is
applied in the time domain and considers only a single characteristic
timescale.  In Maccarone \& Coppi (2002), we adapted the time skewness
statistic of Priedhorsky et al. (1979) by rearranging some terms and
dividing by the cube of the standard deviation to non-dimensionalize
the time skewness.\footnote{We determined empirically in that paper
  that this is a good way to non-dimensionalize the skewness, since
  this approach yields skewnesses in different energy bands, with
  different count rates that are generally quite similar to one
  another as long as the source count rate dominates over the
  background count rate.}

\begin{equation}
TS(\tau) = \frac{1}{\sigma^3}\frac{1}{K} 
\sum_{i=0}^{K-1} (s(t)-\bar{s})^2(s(t-\tau)-\bar{s}) - (s(t)-\bar{s})(s(t-\tau)-\bar{s})^2
\end{equation}

Given that the bispectrum is simply a complex number, it can be thought of as
consisting of a magnitude and a phase.  The phase is called the biphase, and
for reasons that will become clear in the next section, the biphase must be
defined over the full $2\pi$ interval, and not simply as the arctangent of the
imaginary part of the bispectrum divided by the real part of the bispectrum as
is sometimes done in the bispectrum literature.  A version of the magnitude
has been considered most heavily in previous astronomical time series papers.
This version is known as the bicoherence.  It is quite similar to the
cross-coherence used to test whether the time lags between two energy bands
are constant -- it takes on a value from 0 to 1, with 0 indicating that there
is no nonlinear coupling of the phases of the different Fourier components
between different observations, and 1 indicating total coupling.  The most
commonly used expression for the bicoherence is that of Kim \& Powers (1979):

\begin{equation}
b^2(k,l) =
\frac{\left|\sum{X_i(k)X_i(l)X^*_i(k+l)}\right|^2}{\sum{\left|X_i(k)X_i(l)\right|^2}\sum{\left|X_i(k+l)\right|^2}},
\label{bicoeqn}
\end{equation}

where $b^2(k,l)$ is the squared bicoherence -- although see
e.g. Hinich \& Wolinsky (2004) who point out that other methods of
normalization are more sensitive to some types of non-linear behavior.
The Kim \& Powers (1979) normalization has the attractive property
that, for a system with power at only three frequencies, the squared
bicoherence represents the fraction of the power at the third
frequency that can be explained by coupling of the three modes (see
also Elgar \& Guza 1985); such a simple interpretation, however, is
not possible in the cases of broadband coupling (McComas \& Briscoe
1980).

\subsection{Understanding the biphase}
The biphase thus holds powerful information about the shape of the
light curve.  Masuda \& Kuo (1981) worked out some key implications of
the biphase.  In the early papers on the bispectrum (e.g. Hasselmann
et al. 1963), the focus was on the skewness, which is closely related
to the real part of the bispectrum.  Masada \& Kuo's paper examined a
few simple cases, where only three commensurate frequencies were
considered, and present a few examples showing that a positive
skewness when considering a particular set of timescales will manifest
itself in a positive real component for the bispectrum.  The real
component describes the extent to which the flux distribution of the
source is skewed, and the imaginary component described the extent to
which the time series is symmetric in time in a statistical sense.
Poisson noise thus affects only the real component.

\subsubsection{Skewness of the flux distribution}
A positive skewness results from an asymmetric distribution of fluxes,
with a long tail to high flux.  Accreting objects often show
log-normal flux distributions (see e.g. Lyutyj \& Oknyanskij 1987;
Gaskell 2004; Uttley, McHardy \& Vaughan 2005) -- note also that the
fact that the optical fluxes of active galactic nuclei follow a
log-normal distribution is obscurred, to some extent by the use of the
logarithmic magnitude scale, rather than fluxes, for most optical work
(Gaskell 2004).  The log-normal distribution has a positive skewness.
Therefore, any cases of negative real components of bispectra (or,
alternative, of biphases in the range from $\pi/2$ to $3\pi/2$) are
especially interesting, as they indicate particular timescales in
particular observations on which can immediately determine that the
flux distribution is not the ``standard'' log-normal distribution.

The log-normal distribution is often found to provide a good first
order description of the distribution of values of a range of
phenomena, both in nature (e.g. Makuch et al. 1979), and in the social
sciences (the log-normal distribution is an underlying assumption in
the Black \& Scholes 1973 formula for option pricing -- although is
has been argued to underpredict rare events -- e.g. Haug \& Taleb
2011).  As a result, in some cases, it may make sense to apply time
series analysis techniques to the logarithms of the measured values,
rather than to the values themselves -- in such a case finding a
substantial value of the real component of the bispectrum would
indicate that the distribution deviated from log-normal, which might
be more enlightening than demonstrating that the distribution deviates
from being symmetric about the mean.  We do not make calculations of
the properties of the log of the count rate distributions in this
paper, but rather we simply note that it may be worth doing under
certain circumstances (e.g. it may make more sense to work with
magnitudes than fluxes when dealing with bright optical sources, if
the optical flux distribution is expected to be log-normal).

\subsubsection{Asymmetry of the time series}
The asymmetry of the time series is related to the imaginary component
of the bispectrum.  A biphase of $\pm{\pi}/2$ is obtained for a
sawtooth wave.  A sawtooth wave, for example, has a symmetric flux
distribution, and hence a zero real component for the bispectrum and
complete asymmetry in time, and hence a purely imaginary component to
the bispectrum.  The sign convention is such that positive imaginary
components of the bispectrum correspond to sawtooths which rise more
slowly than they fall off.  Again, there are some indications of what
to expect from past measurements.  For X-ray binaries in the hard
state, for example, Maccarone, Coppi \& Poutanen (2000) suggested on
the basis of the combination of hard time lags and narrower
autocorrelation functions at higher energies that the characteristic
variability pattern for these systems must be a relatively slow rise
characterized by a relatively fast fall-off, and that this slow rise
must be slower and start earlier at low energies than it does at
higher energies.  This basic idea, at least for the rapid variability
was verified by use of the time skewness statistic (Maccarone \& Coppi
2002).

We note also that one can also think of the asymmetry in time as being
the skewness of the Hilbert transform of the time series.  The Hilbert
transform, in the Fourier domain, can be executed by shifting the
phases of all positive frequency components by $-90$ degree and all of
the negative frequency components by $+90$ degrees.  Since this
converts sin $x$ to $-{\rm cos} x$, and cos $x$ to sin $x$, we can see
that it bears a relation to the negative derivative of the function
(although no weighting by the frequencies is applied).  This thus
yields some similarity with the time skewness statistic, which is a
flux-weighted average of the slope of the light curve on a particular
timescale.

\section{Plots for extreme cases}
We now present schematic diagrams for a few ``extreme'' cases of light
curves that can occur astrophysically.  These can be used to develop
an intuitive picture of what lightcurves should look like for
different values of the biphase.

\subsection{``Ideal'' pulsars: harmonics with biphase = 0}
First, let us consider an overly simplified description of a pulsar light
curve.  If we suppose that the time series has a series of $\delta-$functions
appearing periodically, where the source is bright, and has zero flux at all
other times, then clearly there is no asymmetry to the time series, but there
is a strong skewness in the flux distribution, with most of the data points at
values much less than the mean.  Such a timeseries will have a positive
skewness on all timescales on which there is power in the power spectrum.  The
biphase will thus generically be 0 wherever there is power.  For pulse shapes
with some asymmetry to them, there will be biphases different from zero, but
always with positive real components to the bispectrum.

\subsection{Eclipses: harmonics with biphase = $\pi$}

\begin{figure*}
\includegraphics[width=2.5 in,
  angle=-90]{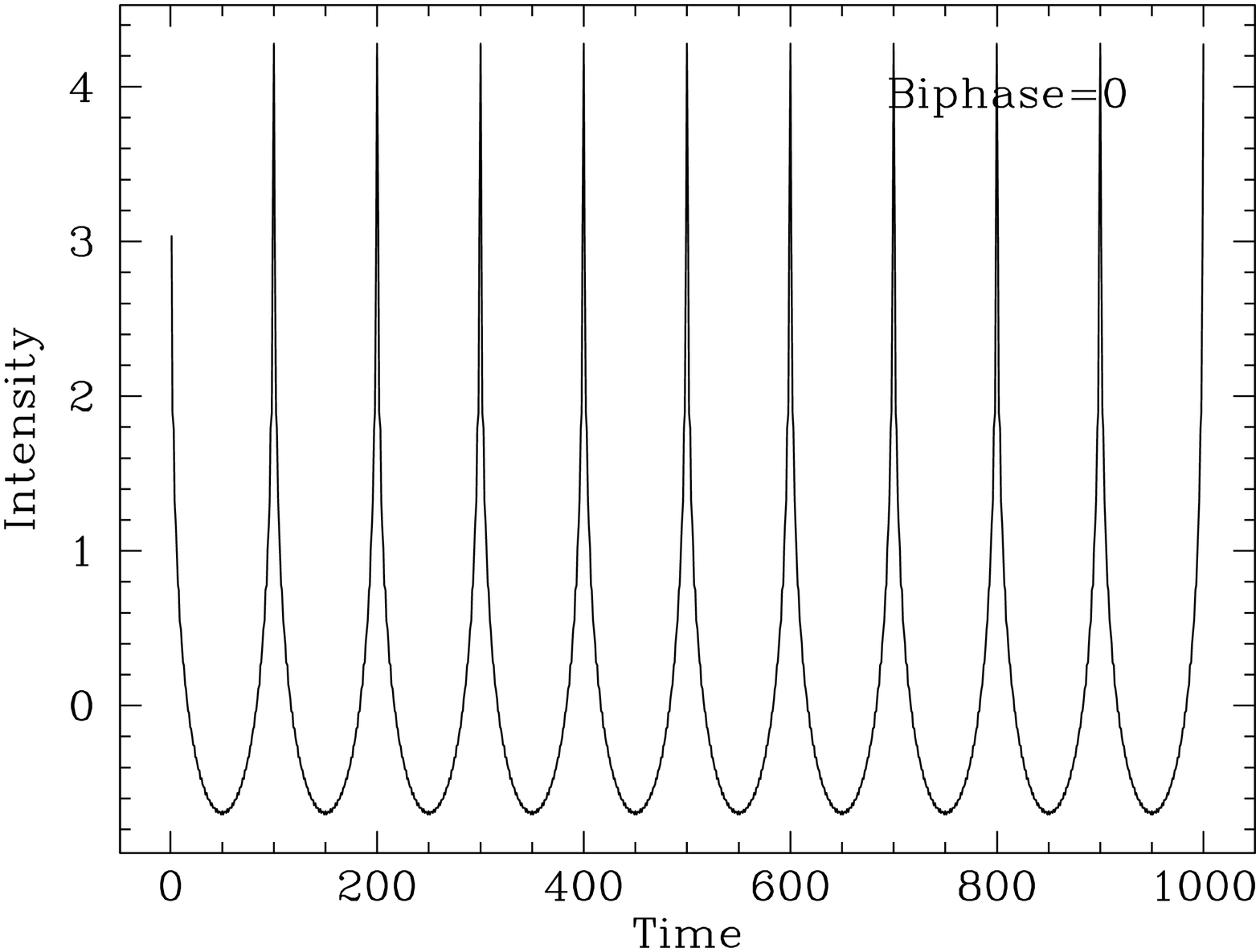}\includegraphics[width=2.5 in,
  angle=-90]{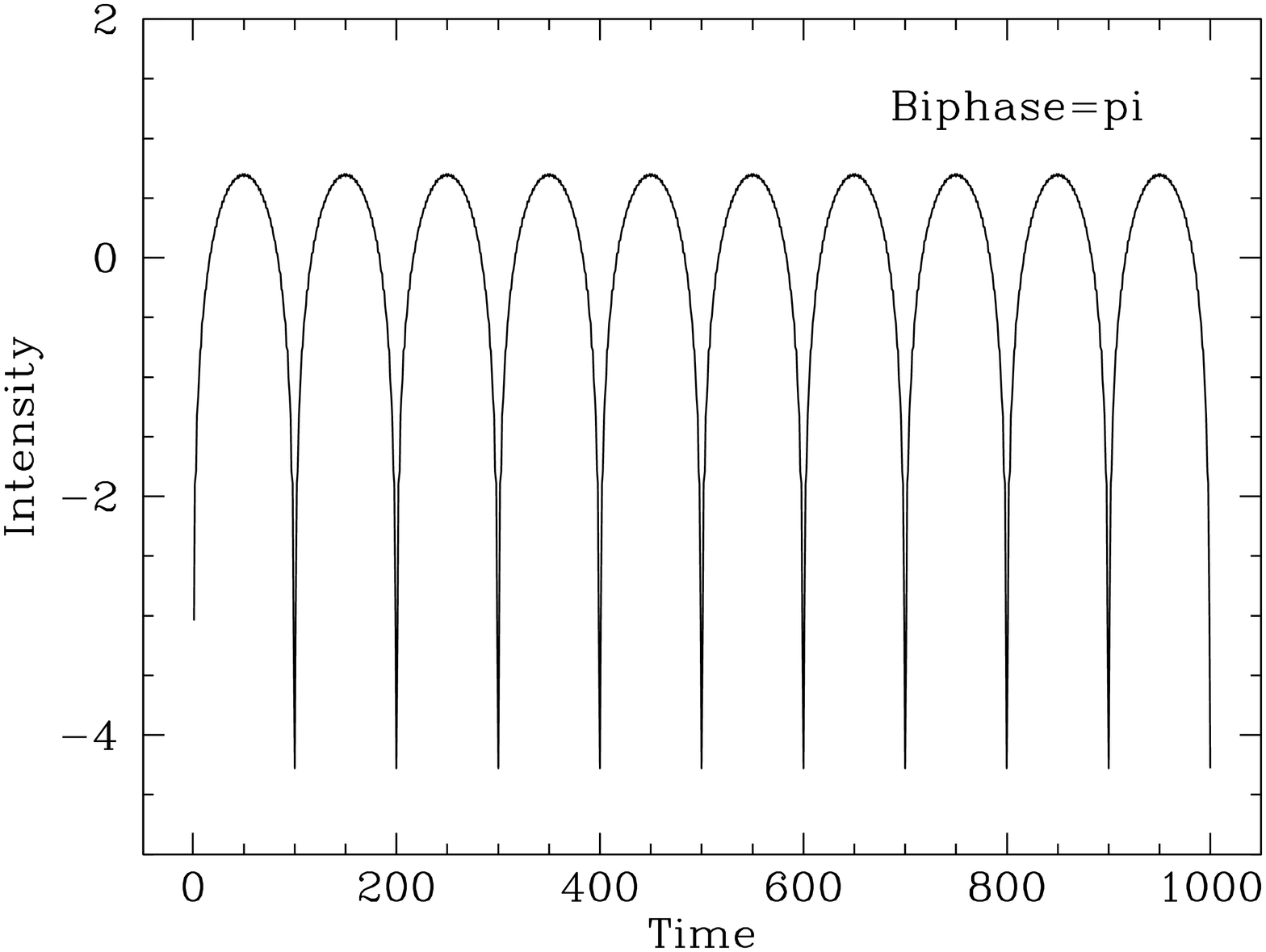}\\
\caption{Left: a pulsar-like time series, produced by adding cosines
  all with phase 0, frequencies of $j{\omega_0}$, where $\omega_0$ is
  the fundamental frequency of the oscillation, and $j$ are all
  integers.  Only the first 40 terms of the sum are added, and the
  normalization of each component is taken to be $1/j$.  Right: an
  eclipser-like time series, produced in the same manner as the
  pulsar-like time series, except that a phase of $\pi$ is added to
  each cosine, giving the triplets of harmonics biphases of $-\pi$.}
\label{biphasezeropi}
\end{figure*}

Next, we can consider the opposite case: periodic drops in flux of a constant
amount which occur periodically.  Apart from a DC offset and a constant of
proportionality, this scenario is essentially the same as taking each flux
value to be the negative of the flux values of the scenario above.  This gives
a strong negative skewness, and no asymmetry.  The biphase should thus be
$\pi$ for all cases where there is any Fourier power.  The plots of time series which have biphases of 0 and $\pi$ are presented in figure \ref{biphasezeropi}.

\subsection{Sawtooth oscillations: harmonics with biphase of $\pm\pi/2$}

While sawtooth oscillations are not common in astrophysics, there are a few
cases where they are seen.  For example, solar flare radio emission can
sometimes show linear rises followed by rapid drops in flux (Klassen et
al. 2001) -- and these sawtooth oscillations are common in other kinds of
magnetic reconnection scenarios (Zweibel \& Yamada 2009).  Classical Cepheids
and RR Lyrae stars show the opposite - sharp rises in flux, followed by slow,
linear decays.  Rapidly rising, linearly fading sawtooths will give $-\pi/2$
for the biphase and linearly rising, rapidly fading sawtooths will give
$\pi/2$ for the biphase.  Generically, any function which rises more sharply
than it fades will have a negative imaginary component of the biphases and and
function which fades more sharply than it rises will have a positive imaginary
component of the biphase.

\begin{figure}
\includegraphics[width=3.5 in]{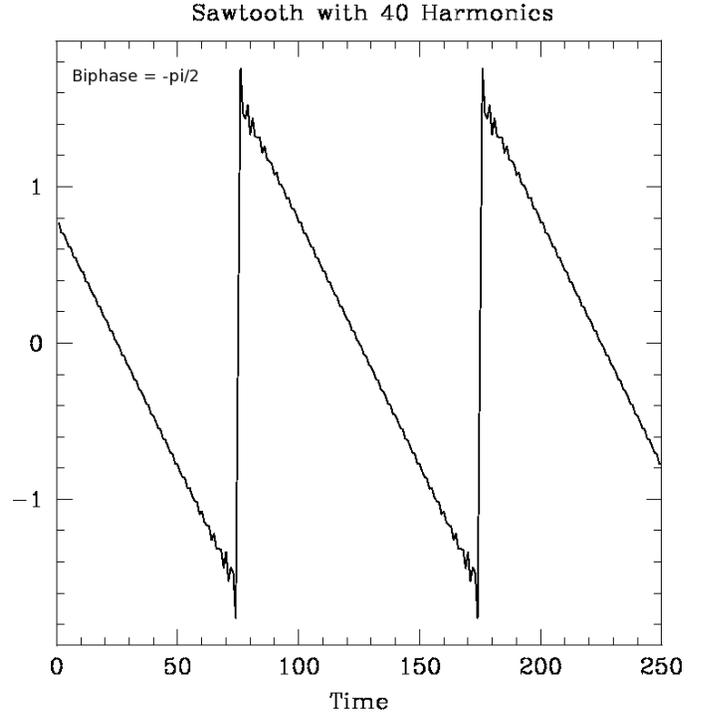}
\caption{This is a summation of the first 40 harmonics for a sawtooth
  oscillation with a biphase of $-\pi/2$.}
\label{sawtooth}
\end{figure}

A strict sawtooth wave is defined as the summation over all integer
values of $j$ of $\frac{1}{j}{\rm cos} [j \omega_0 t +
  (j-1)\times{\pi}/2]$.  The constant added phase may alternatively be
multiplied by $-1$ to allow the opposite sense of symmetry.  We plot
the summation of the first 40 terms of a sawtooth oscillation in
figure \ref{sawtooth}.  As the number of terms approaches infinity,
the wave form approaches a strict instantaneous rise, linear decay (or
linear rise, instantaneous decay).  

\subsection{Beyond the simple examples}

It is important to remember, also that the biphase alone does not determine
the shape of a time series.  The examples listed above are the easiest cases
to visualize for their particular values of the biphase.  However, these are
all cases where the power spectrum is composed solely of a fundamental
frequency and its overtones, and where the amplitudes of the different
harmonics are set in a specific manner.  In a sawtooth wave, for example, the
normalization of the sine wave corresponding to a particular harmonic is
inversely proportional to its frequency.  A system with the same set of
biphases, but a different power spectrum, would similarly be strongly
asymmetric in time, but could have a light curve with a qualitatively
different shape.  

\begin{figure}
\includegraphics[width=3.5 in]{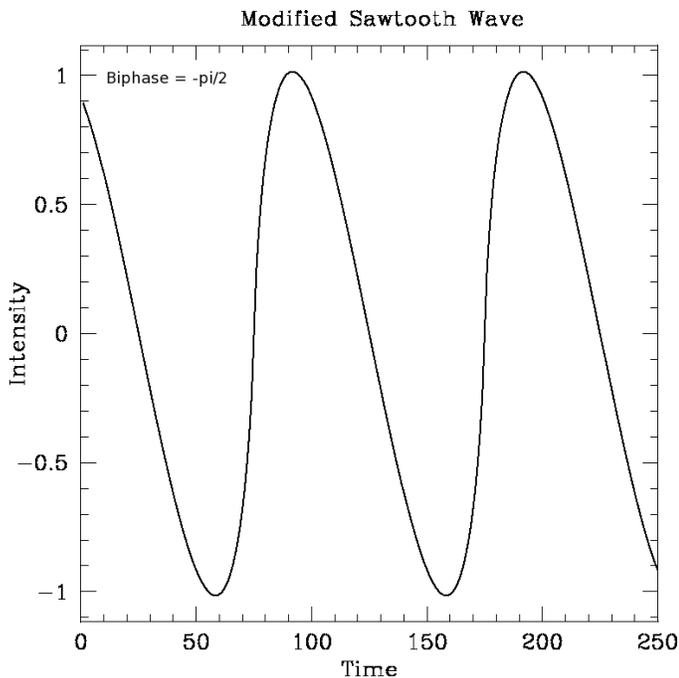}
\caption{This is a modified sawtooth.  The frequencies which contribute, and
  the phases at those frequencies are the same as for the sawtooth wave in
  figure \ref{sawtooth}, but the amplitudes of the cosine waves at the
  different harmonics have been changed to scale as $1/j^2$ instead of scaling
  as $1/j$.  This sawtooth wave has biphase $-\pi/2$ for all combinations of
  harmonics in which the two lower frequencies add up to the higher
  frequency.}
\label{mod_saw}
\end{figure}

For example, in figure \ref{mod_saw} we plot a modification of the sawtooth
wave.  Like in figure \ref{sawtooth}, we plot the sum of 40 harmonics, all as
cosine waves with an added phase of $\pi/2 \times (j-1)$, for the $j$th
harmonic.  Instead of using a normalization of $1/j$ for the Fourier spectrum,
we use $1/j^2$, which gives more weight to the lower harmonics, and thus
results in the more curved shape to the time series near the peak.  The time
series is still symmetric in its flux distribution, and asymmetric in time,
and so it still has biphase of $-\pi/2$.

\subsection{Beyond a pure harmonic structure}

The easiest examples for which to attempt to visualize the bispectrum are the
cases where all the relevant frequencies are integer multiples of a
fundamental frequency.  The real power spectra of many interesting classes of
astrophysical objects, including, but not limited to, accreting compact
objects with low magnetic fields, have quite broad power spectra.  At the
present time, it has been shown only for GRS~1915+105 that this broad power
spectrum carries a nonlinear relationship with the quasi-periodic oscillations
seen in the source (Maccarone et al. 2011).  

Given this finding, it is of interest to show what the light curves
will look like for different types of power spectra and different
values of the biphase, in order to help develop an intuition for the
meaning of the biphase.  We thus present some calculations of
simulated light curves for such Fourier spectra.  We generate a
Fourier spectrum using an approach similar to that taken in Timmer \&
K\"onig (1995) -- see also Davies \& Harte (1987) -- with some small
modifications.

First, we consider examples where the two noise components are both at
lower frequencies than the QPO frequency.  We draw a random amplitude
at each frequency such that the power spectrum will take a value
uniformly distributed between 0 and 2 times the desired power spectrum
level at that frequency.  We draw phases randomly for the lower
non-QPO frequency, and then force the combination of the higher
non-QPO frequency's phase and the QPO frequency's phase to give the
biphase at the desired value -- this forces the biphase to have a
particular value for the case where $f_1+f_2=f_{QPO}$.  There is then
no consistent value of the biphase for the coupling within the noise
component, but there is a consistent value of the biphase of the QPO.

\begin{figure*}
\includegraphics[width=2.5 in,
  angle=-90]{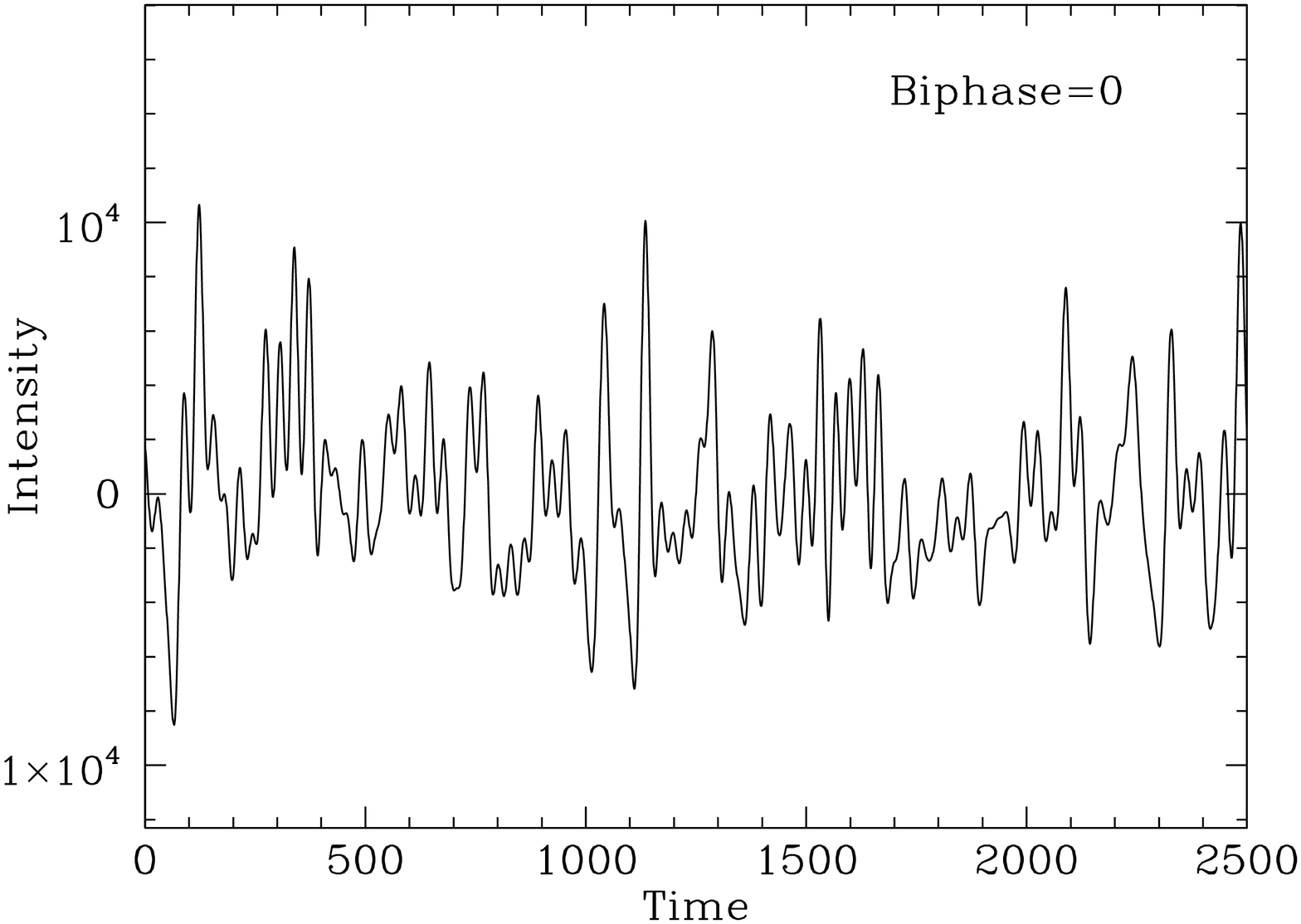}\includegraphics[width=2.5 in,
  angle=-90]{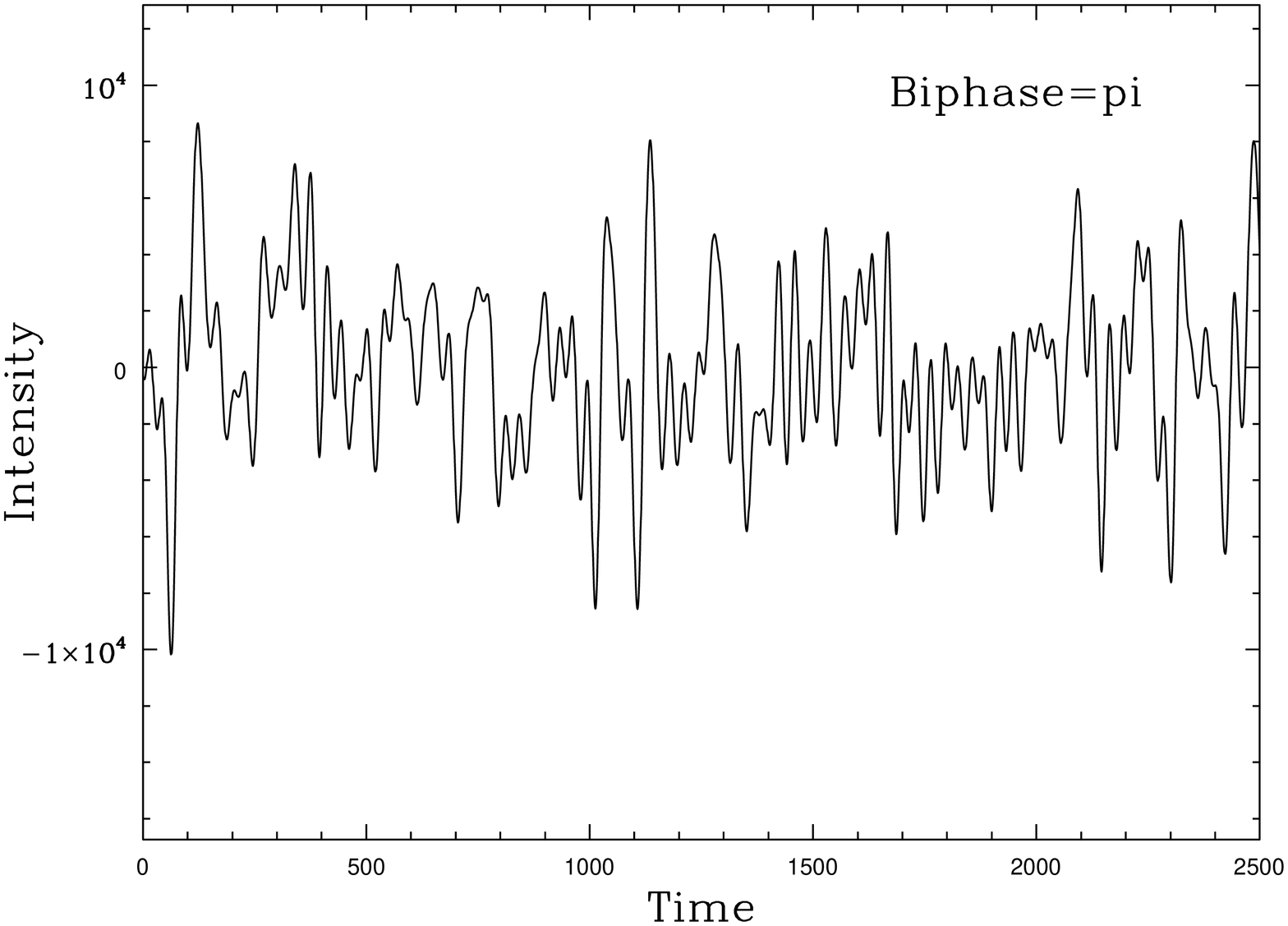}\\
\includegraphics[width=2.5 in, angle=-90]{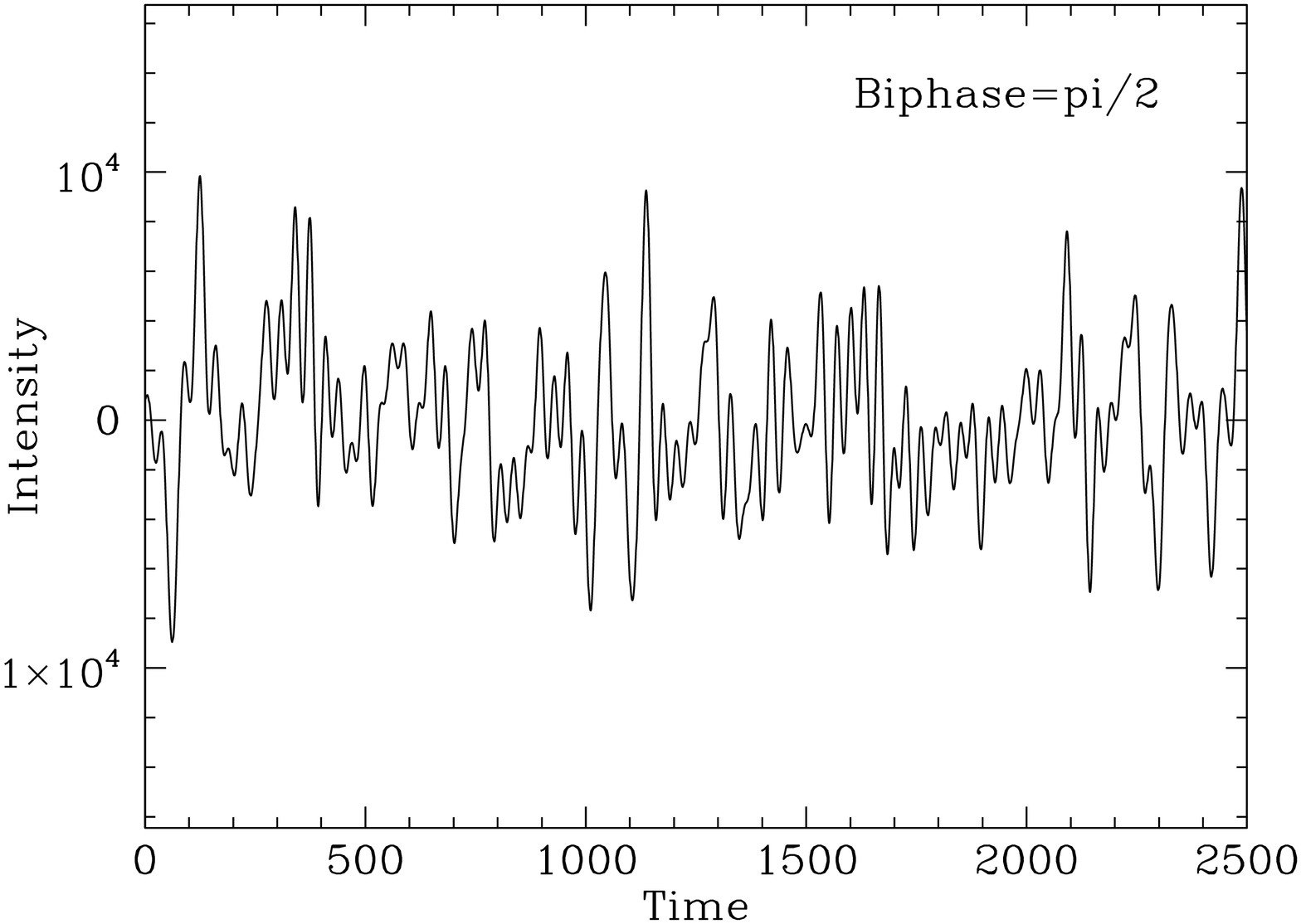}\includegraphics[width=2.5in,    angle=-90]{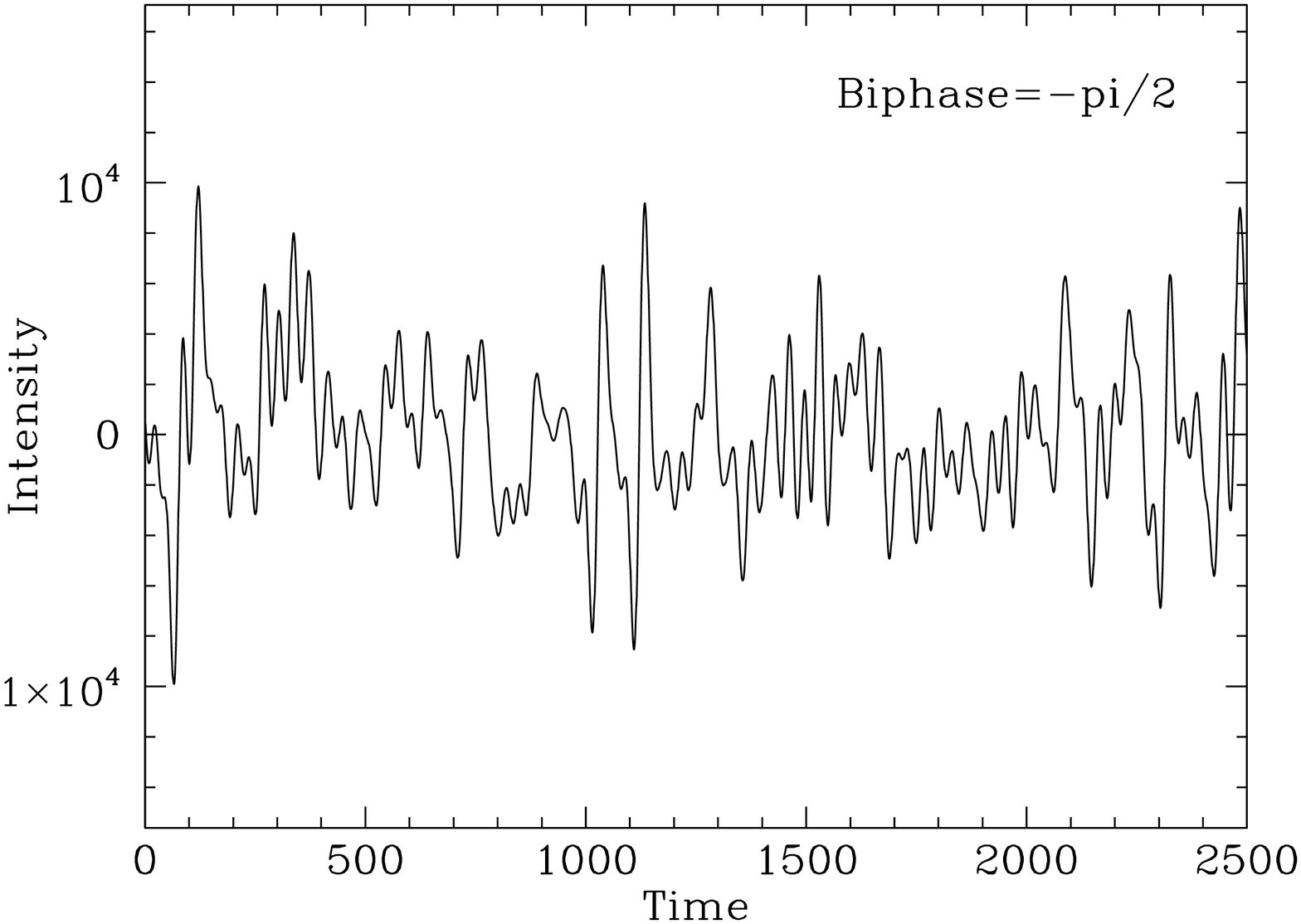}\\
\caption{The bispectra for the four cases where all the noise power is at
  lower frequencies than the QPO power.  The figure in the upper left has a
  biphase of 0, the figure in the upper right has a biphase of $\pi$, the
  figure in the lower left has a biphase of $\pi/2$ and the figure in the
  lower right has a biphase of $-\pi/2$.  The same seed was used for the
  random number generator for all four cases, resulting in some features
  dominated by the noise component looking similar in all four plots.}  
\label{lfnoise_examples}
\end{figure*}

\begin{figure*}
\includegraphics[width=2.5 in,
  angle=-90]{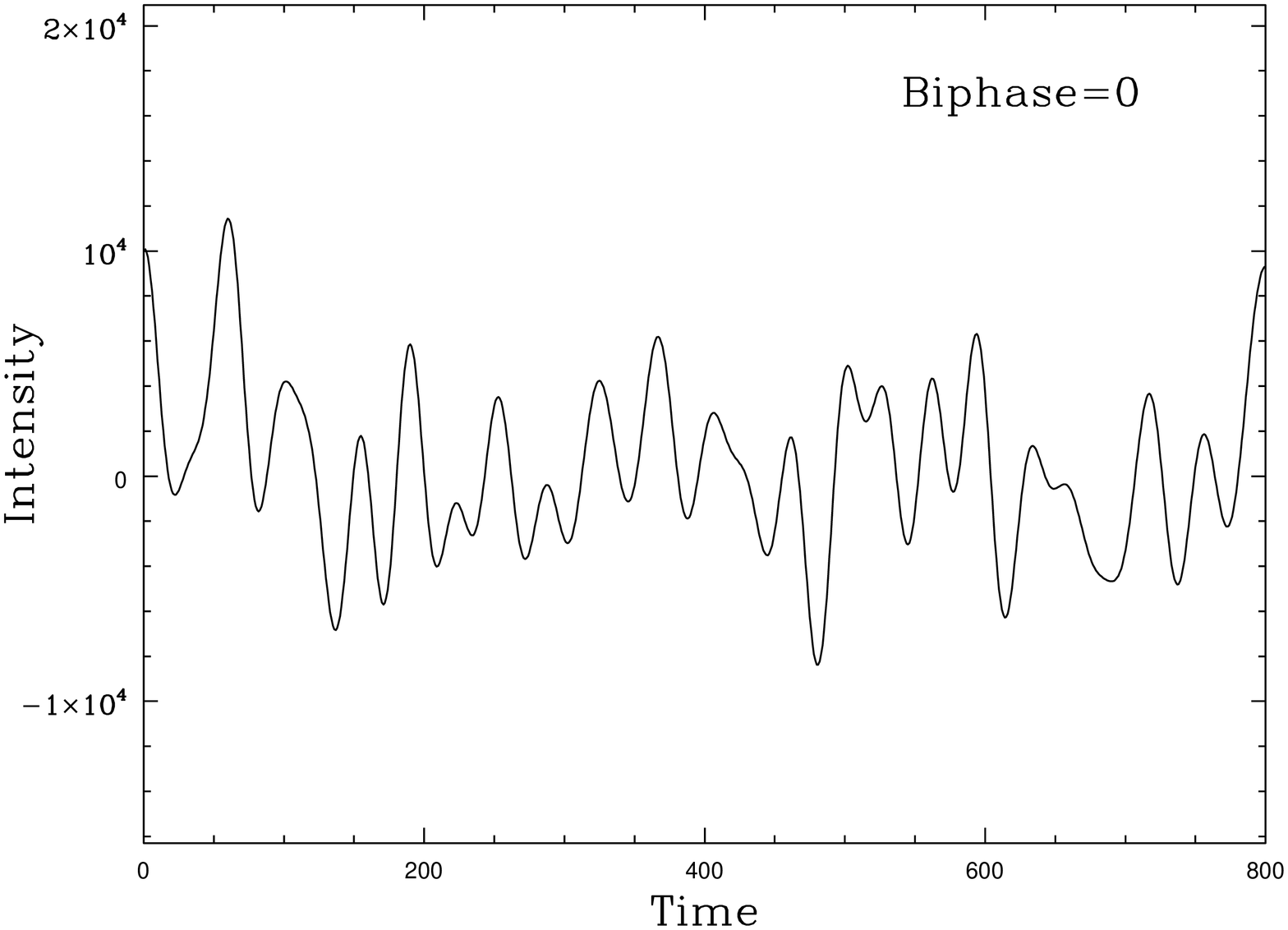}\includegraphics[width=2.5 in,
  angle=-90]{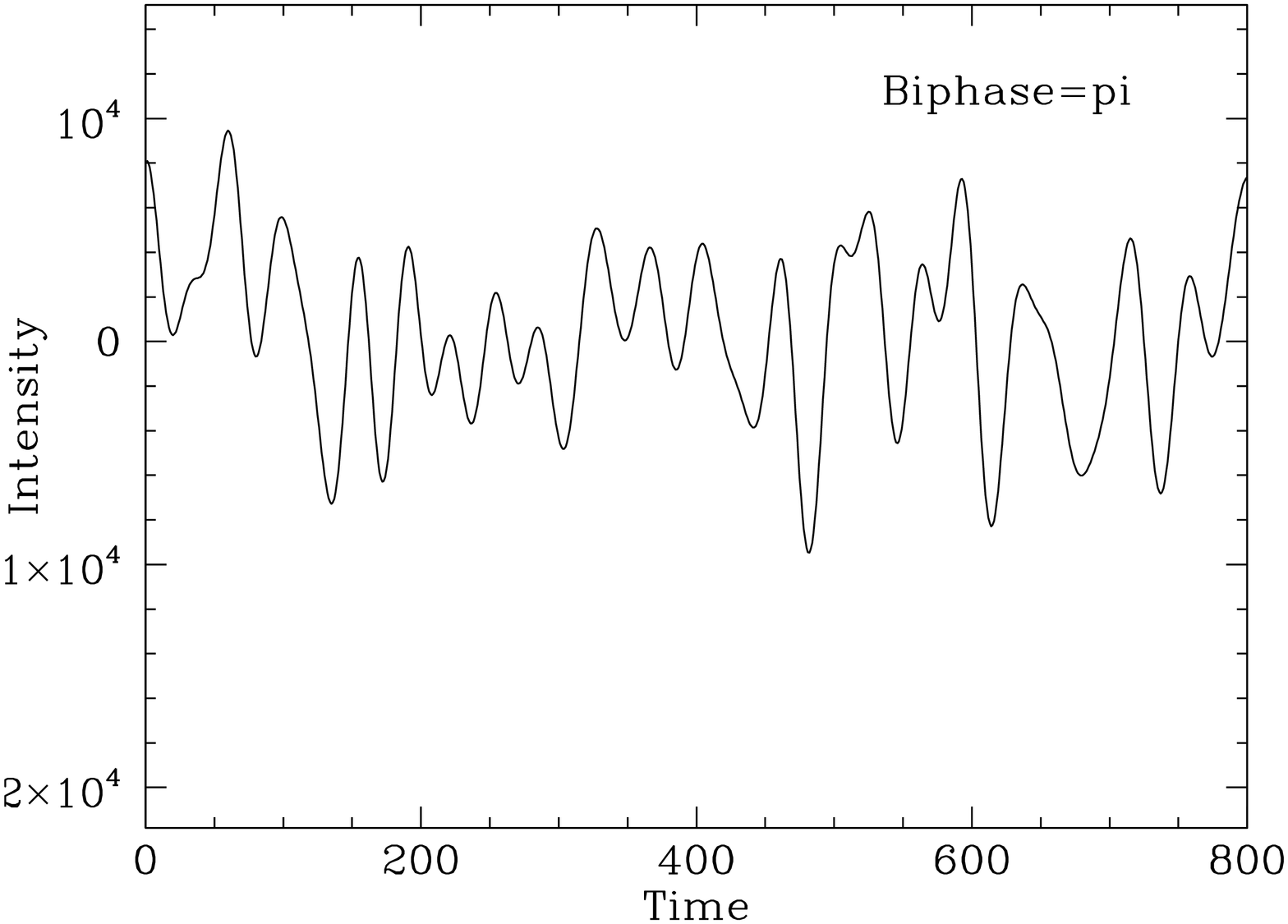}\\
\includegraphics[width=2.5 in, angle=-90]{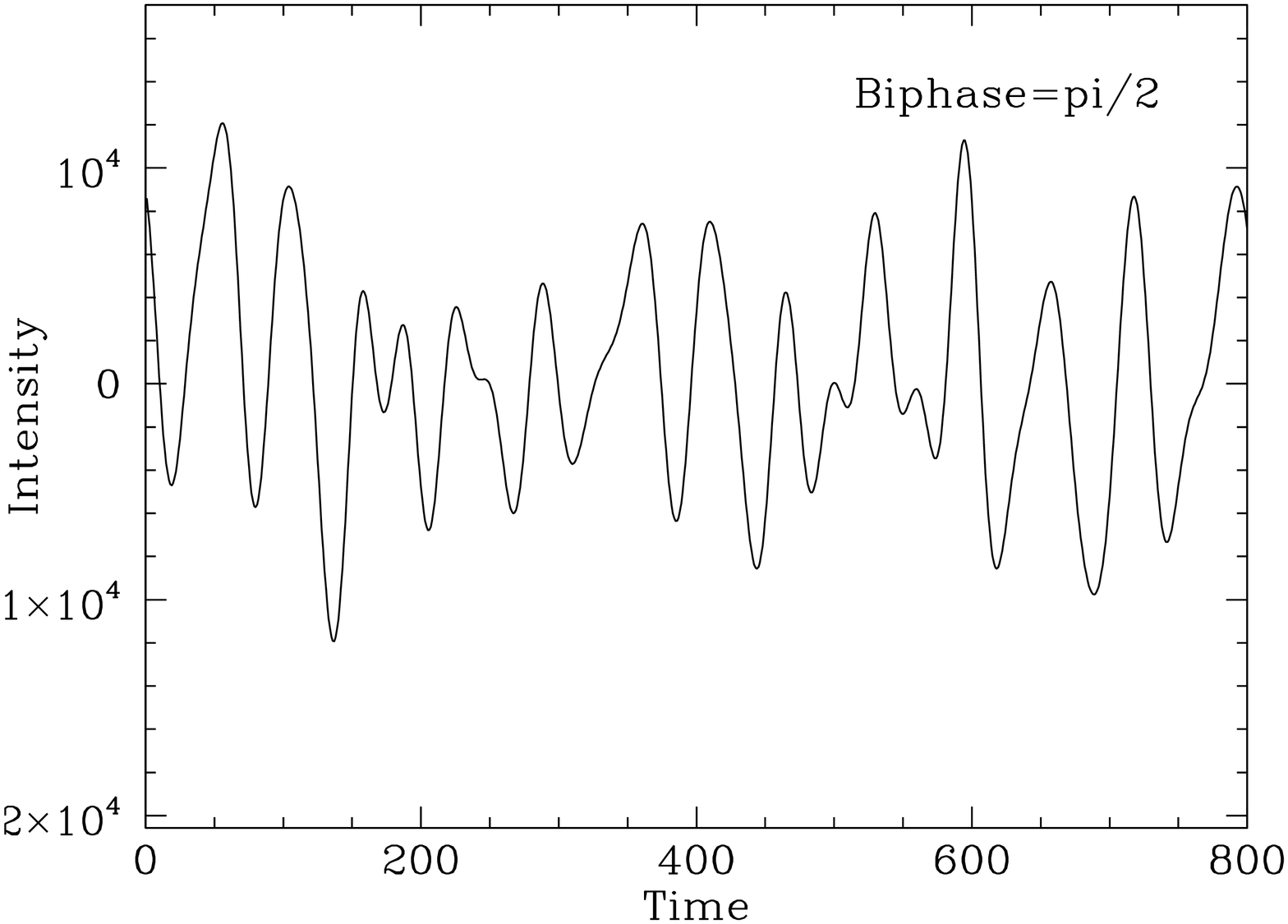}\includegraphics[width=2.5in,    angle=-90]{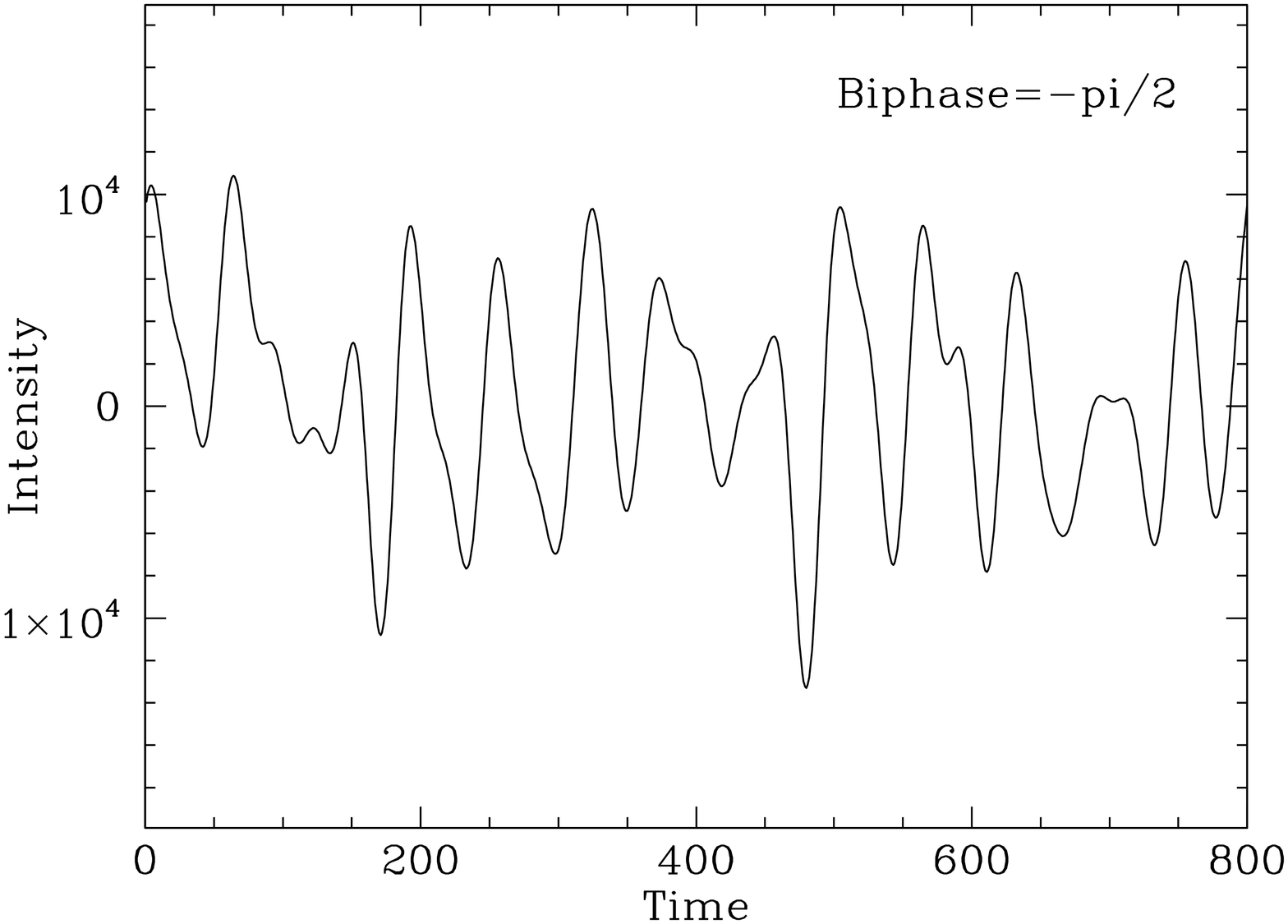}\\
\caption{The bispectra for the four cases where QPO frequency is below
  the noise break frequency.  The figure in the upper left has a
  biphase of 0, the figure in the upper right has a biphase of $\pi$,
  the figure in the lower left has a biphase of $\pi/2$ and the figure
  in the lower right has a biphase of $-\pi/2$.  The same seed was
  used for the random number generator for all four cases, resulting
  in some features dominated by the noise component looking similar in
  all four plots.}
\label{lfqpo_examples}
\end{figure*}

We compute some illustrative examples of time series with different
values of the biphase.  We treat the noise as a power spectrum with a
broken power law, with flat power below a break frequency, and a
$\nu^{-1}$ slope above the break frequency.  The break frequency is
set to be 1/4 of the QPO frequency.  The QPO is modelled as a peak at
a single frequency (i.e. it is taken to be strictly periodic) for the
sake of simplicity.  Above the QPO frequency, we assume there is no
Fourier power.  We consider the cases with biphases of 0, $\pi/2$,
$\pi$, and $-\pi/2$, and plot them in figure \ref{lfnoise_examples}.
We then perform the same procedure as above, except for a case where
the QPO frequency is set to be about 2/3 of the break frequency of the
noise power spectrum.  The results of these calculations are plotted
in figure \ref{lfqpo_examples}.

The basic structures of the signals still agree with the idealized
cases discussed above.  When the biphase is 0, one can see that the
deviations from the mean are sharper, but less frequent in the
positive direction than in the negative direction, while the opposite
is true for biphase of $\pi$.  Furthermore, computation of the
skewness of the time series yields positive values for the former case
and negative values for the latter case.  When the biphase is $-\pi/2$,
one can see that there are sharp rises from the mean, followed by
slower decays in the value of the time series, and the oppsite is true
for a biphase of $\pi/2$.

\subsection{Nonlinear coupling with zero bicoherence}
There {\it are} cases where coupling clearly exists between different
frequencies in a time series, but where the bispectrum will tend to
zero.  An example is a square wave.  Square waves have symmetric flux
distribution and are symmetric in time.  Triangle waves lack a
bispectrum, but have nonlinear couplings in the same manner.  These
can both be understood in a straightforward manner by considering the
harmonics that contribute to the waves.  Both triangle waves and
square waves are composed of the sums of odd integer harmonics only.
Since no two odd numbers can sum to make a third odd number, it is not
possible to find sets of frequencies in square waves or in triangle
waves for which the bispectrum will take a non-zero value.

Still higher order polyspectra can, in principle, be used to
investigate the even higher moments of light curves that can
characterize such waveforms.  At the present, we stick with bispectral
analysis because even it is not yet well understood, at least in
astronomy, and because the statistical errors on the trispectrum, the
next higher moment, are likely to be too large to make good use of
that statistics with existent data sets.\footnote{We note that we have
  not attempted to make any computations of the trispectrum yet, and
  that our pessimism may be unwarranted, so we do not wish to be
  overly discouraging to others trying to compute it with RXTE data.}

In the near term, one could consider simply constructing flux
distributions and attempting to determine whether the kurtosis of the
flux distribution differs strongly from the expectations for a pure
log-normal distribution.  If so, then the trispectrum would be worth
computing to try to isolate the timescales contributing most strongly
to the kurtosis.  If not, then the trispectrum may still contain
information not contained within the power spectrum and the
bispectrum, but the computational difficulties in computing the
trispectrum, along with the difficulties in visualizing and
interpreting it for non-harmonic frequencies may make the exercise
unproductive.

The LOFT mission (Feroci et al. 2012) should open up the possibility
of computing trispectra in cases where RXTE can be used only to
compute bispectra.  With about 20 times the collecting area, it should
be possible to deal with an additional term with associated
uncertainties in the numerator of the polyspectrum, but a set of
simulations and discussion of this topic goes well beyond the scope of
this paper.  The difficulties in interpreting the trispectrum would
remain, but an approach like that in this paper, considering first
some simple cases like square waves and triangle waves, could be used
as a starting point if some statistically significant signals were
detected.

\section{An application outside high energy astrophysics: exoplanet searches}

One of the key sources of background for exoplanet transit detections is
stellar activity (e.g. Aigrain, Favata \& Gilmore 2004). Planetary transits
should fit the following characteristics:
\begin{enumerate}
\item A time-symmetric time series (apart from some very weak effects due to the
  rotation of the star being eclipsed -- Rossiter-McLaughlin)
\item Strong harmonic structure, with the relative intensities of the different harmonics depending on the duration of the eclipse.
\item Biphase of $\pi$.
\end{enumerate}

The bispectrum may then hold the possibility to allow the detection of
moderate strength transits even in stars with strong noise.  While the
flickering from a sea of weak stellar variability may make it difficult to
prove, from a power spectrum alone, that a particular star is transiting a
planet, the bispectrum may in some cases indicate that particular frequencies
show the coupling expected for an eclipse.  An added benefit is that the soft
X-rays produced by stars with strong activity are strongly absorbed by oxygen,
so there might be hope to search for oxygen in these stars' atmospheres more
readily than can be done in the optical.  In cases of elliptical orbits where
effects such as Doppler beaming, ellipsoidal modulations, and reflection may
have asymmetric modulations on the orbital period, the expectation value of
the biphase may not be strictly $\pi$ -- but in general, the amplitudes of such
variations will be a few orders of magnitude less than the amplitude of
variations due to transiting (Loeb \& Gaudi 2003).  In fact, it may be more
likely that the bispectrum can be used to help establish the nature of
periodicities due to non-eclipsing planets than that it will hinder the use of
the bispectrum for detecting planets.

The passages of star spots may also produce bispectra with biphases
near $\pi$.  A key difference is that the effects of star spots are
not consistent from orbit to orbit.  The periods change as the spots
move away from the equator, and even at constant lattitude, the phases
change as spots are created and destroyed.  As a result, the
bicoherence may be a good means of separating starspot activity from
transiting.

\section{Application to the biphase of GRS~1915+105}

In a previous paper (Maccarone et al. 2011 -- M11), we showed that the
X-ray binary GRS~1915+105 shows strong bicoherence in interactions
between its strong quasi-periodic oscillations and its broadband
noise.  For this system, several different patterns of variability
were seen in the bicoherence plots in the data.  The biphase of that
system was not, however, considered in our previous paper.  We use the
same computations of the bispectrum from exactly the same data sets we
used in M11.  Those computations were made by taking long observations
of GRS~1915+105 during which the power spectrum appeared stationary,
taking a series of Fourier transforms, and then following equation
\ref{bispeceqn} and equation \ref{bicoeqn} above.  In the process of
preparing this paper for publication, we realized a clerical error in
M11 resulted in the wrong time resolution being given in the text for
the time resolution used for observation 10408-01-25-00 -- it was
1/128 second, rather than 1/64 second.

\subsection{A review of the bicoherence results}

If one considers the two lower frequencies involved in coupling, with
the understanding that the third frequency follows trivially from the
first two, then we can consider which values of $f_1,f_2$ show strong
coupling with $f_1+f_2$.  The three characteristic patterns found were
called the ``web'', ``cross'' and ``hypotenuse'' patterns.

The web pattern is characterized by strong bicoherence for $f_1+f_2$ =
$f_{QPO}$ where $f_{QPO}$ is the strongest quasi-periodic oscillation in the
power spectrum; for $f_1=f_2=f_{QPO}$, and for $f_1=f_2=2f_{QPO}$; and is
weaker, but still clearly detectable for $f_1=2f_{QPO}, f_2>2f_{QPO}$.  The
hypotenuse pattern is quite similar in appearance to the web pattern, except
that only the $f_1+f_2$ = $f_{QPO}$ and the harmonic are seen strongly.  The
cross pattern shows strong bicoherence whenever the QPO frequency is either
$f_1$ or $f_2$, but not for the case $f_1+f_2=f_{QPO}$.

\subsection{RXTE observation 10408-01-25-00: a low frequency QPO with a ``web'' pattern bicoherence: dips in the light curve for the fundamental plus second harmonic}

\begin{figure*}
\includegraphics[width=6 in, angle=0]{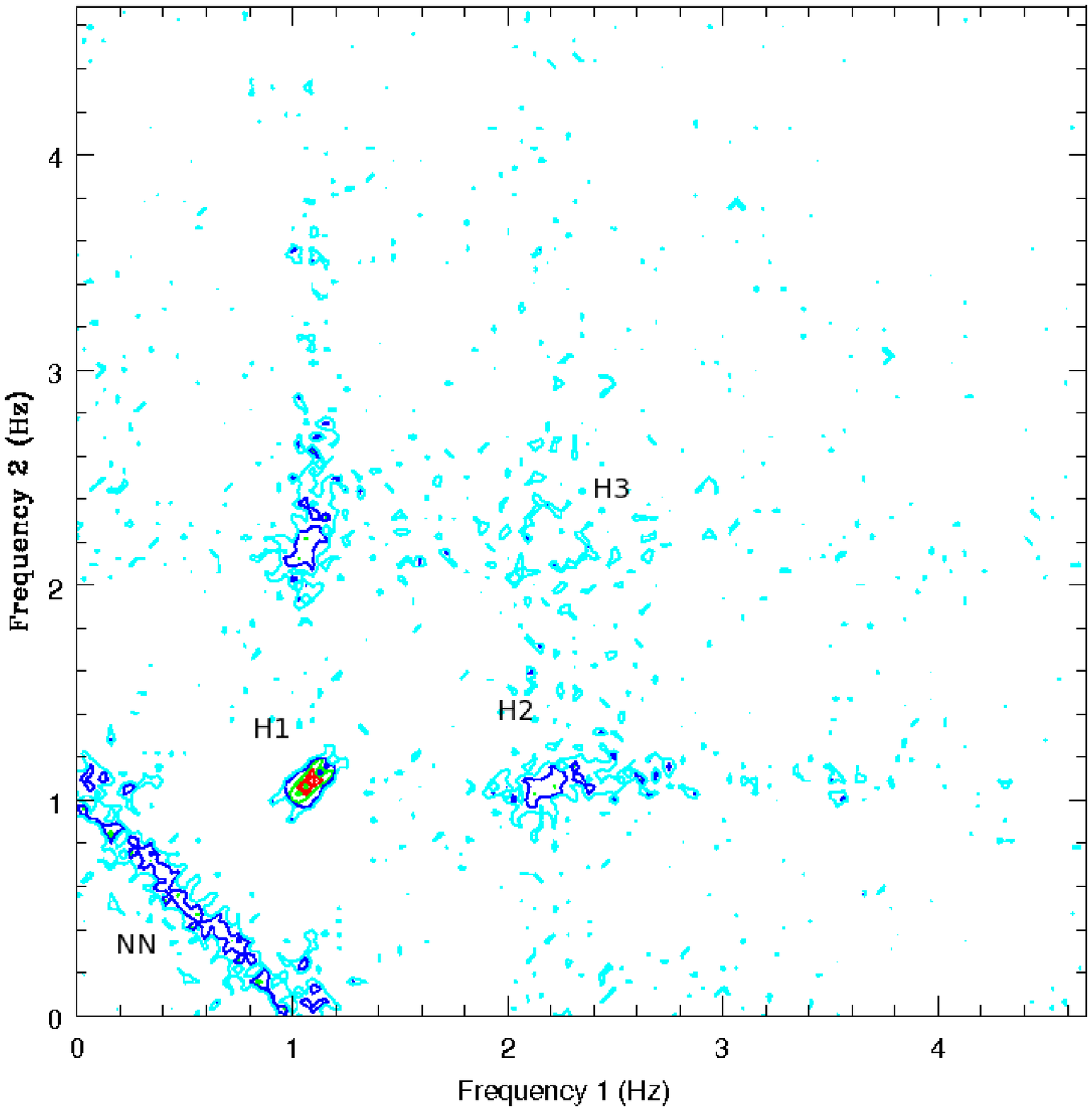}
\caption{The bicoherence plot for observation 10408-01-25-00.  The
  peak closest to H1 is the interaction between the fundamental and
  the second harmonic, the peak closest to H2 is the region for the
  interaction between the fundamental and the second harmonic to
  produce the third harmonic, and the weak peak closest to H3 is the
  interaction of the second harmonic with itself to produce the fourth
  harmonic.  The diagonal stripe marked NN is the region where the two
  noise components add up to the frequency of the fundamental.}
\label{bico1}
\end{figure*}

First, we present the bicoherence plot for observation 10408-01-25-00,
in figure \ref{bico1}.  We have modified the plot from the version
presented in M11, so that the regions we discuss in the text here can
be more readily identified.  Our aim in this paper is not to discuss
the strength of the bicoherence, but merely to show that we are
calculating the biphase for specific regions of strong bicoherence.

In this system we can examine the biphases of the few strongest peaks
in the bispectrum to get a feel for the properties of the strongest
set of waves in the system.  The peak of our bicoherence is $f_1=36,
f_2=37$ with the units of the frequency here being the frequency
resolution of the power spectrum, 0.03125 Hz (i.e. 1/32 Hz).  The
biphases for all combinations where $f_1$ and $f_2$ both range from 35
to 38 times the frequency resolution are found in the range from
0.82$\pi$ to $0.97\pi$, with the mean being $0.86\pi$ and the standard
devation on the phase, estimated from the variance in the measured
values, is $0.05\pi$ -- this can be seen in the bicoherence plot as
the region marked H1 in figure \ref{bico1}.  For the case where the
first and second harmonics interact to produce the third harmonic,
with $f_1$ from bins 35 to 38 and $f_2$ form bins 71 to 75, then mean
value of the biphase is $0.22\pi$ with a standard deviation of
$0.08\pi$ and a standard deviation of the mean of $0.02\pi$ -- this
can be seen as region H2 in figure \ref{bico1}.  For the case where
the second harmonic interacts with itself to produce the fourth
harmonic, (i.e where $f_1$ and $f_2$ range from 71 to 75 times the
frequency resolution of the power spectrum), we find that the mean
biphase is $0.11\pi$, with a standard deviation of $0.17\pi$ and a
standard deviation of the mean of $0.04\pi$ -- this can be seen as
region H3 in figure \ref{bico1}, and it is clear from this figure that
the bicoherence is quite weak at this peak.  Because the fundamental
and the second harmonic are stronger than the third and fourth
harmonics, and the bicoherence is also stronger for the first two
frequencies, the effects of the first two frequencies' phase couplings
have the major impact on the shape of the QPO in the light curve.

\begin{figure}
\includegraphics[angle=-90,width=3 in]{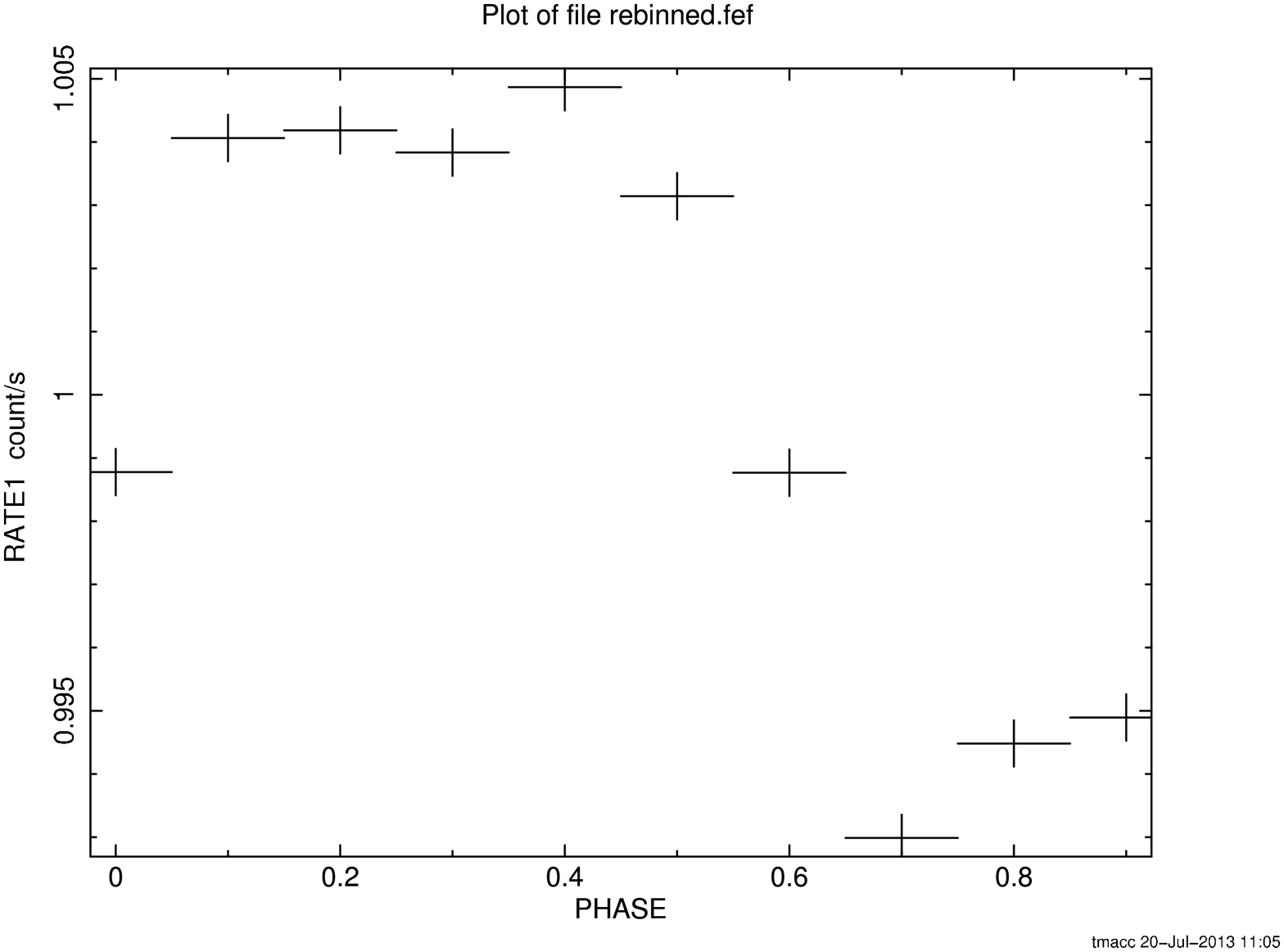}
\caption{The light curve for observation 10408-01-25-00 folded on the period
  of the QPO of 0.849 seconds.  One can clearly see the flux skewness in the
  time series.}
\label{folded10408}
\end{figure}

The biphase 0.86$\pi$ is quite close to $\pi$ itself.  This means that
the shape of the oscillation, as determined by just the two strongest
frequencies is one marked by a relatively smooth profile apart from
deep dips occurring on the fundamental frequency, with phase width of
less than $\pi$.  The folded light curve of the data, made using the
\texttt{efold} tool within FTOOLS, shown in figure \ref{folded10408}
looks as expected.\footnote{The other observations also show folded
  light curves consistent with the qualitative expectations based on
  the biphases, but we do not plot them in the interests of keeping
  the paper from becoming too long.}

We can also consider the case where the two noise frequencies add up
to the fundamental QPO frequency, in the region marked NN in figure
\ref{bico}.  This, in fact, is the real power of the bispectrum, as
this information cannot be obtained by folding the data on the QPO
period.  Here, we take call bispectrum measurements where $f_1+f_2$
ranges from 35 to 38 in units of the frequency resolution.  We find
all the values of the biphase to be between -0.12$\pi$ and 0.33$\pi$
with a mean value of the biphase of 0.14$\pi$, with a standard
deviation of 0.09$\pi$ and a standard deviation of the mean of
0.01$\pi$.  The value of this biphase is then a bit less than
$\frac{1}{7}\pi$ -- i.e. it is much closer to zero than to $\pi/2$,
and the major property of the behavior of the source in terms of the
interactions of the noise and the QPO should be that those expected
for a ``pulsar''-like system.  That is, the ``envelope'' on which the
QPO is imposed should be one of spikes shooting well above a baseline
flux which is slightly below the mean.  This is, in fact the case.
When the light curve is binned on a timescale of $\frac{1}{16}$
second, the mean count rate is 8164 cts/sec, the minimum is 4160
counts/sec, and the maximum is 16272 counts/sec, roughly a factor of
two above and below the mean.  The skewness of the flux distribution
is positive, as calculated using \texttt{lcstats} from within the
FTOOLS.  The system thus largely follows the expectation for a
lognormal flux distribution, as is often observed for X-ray binaries'
light curves (Uttley et al. 2005).  The deviation from a biphase of
exactly zero suggests that on long timescales, the intensity of the
QPO rises sharply and falls more slowly.  While some power is present
in the bicoherence above the QPO frequency, this power is weak, so we
do not investigate the biphase there.

\subsection{RXTE observation 20402-01-15-00: a medium frequency QPO with a ``cross'' pattern bicoherence: a nearly sawtooth pattern for the fundamental plus second harmonic}

In observation 20402-01-15-00, the ``cross'' pattern is seen -- the
bicoherence is strong when the QPO has either the median or the lowest
frequency of the three frequencies being considered, but the bicoherence is
not above the noise level for the case where $f_1+f_2=f_{QPO}$, and $f_1,f_2$
are two noise frequencies.  We can now look at the biphases of the source.
The QPO here peaks in frequency bin 72 (corresponding to a frequency of 2.2
Hz, given the 1/32 Hz frequency resolution), and shows a strong second
harmonic.

\begin{figure*}
\includegraphics[width=6 in, angle=0]{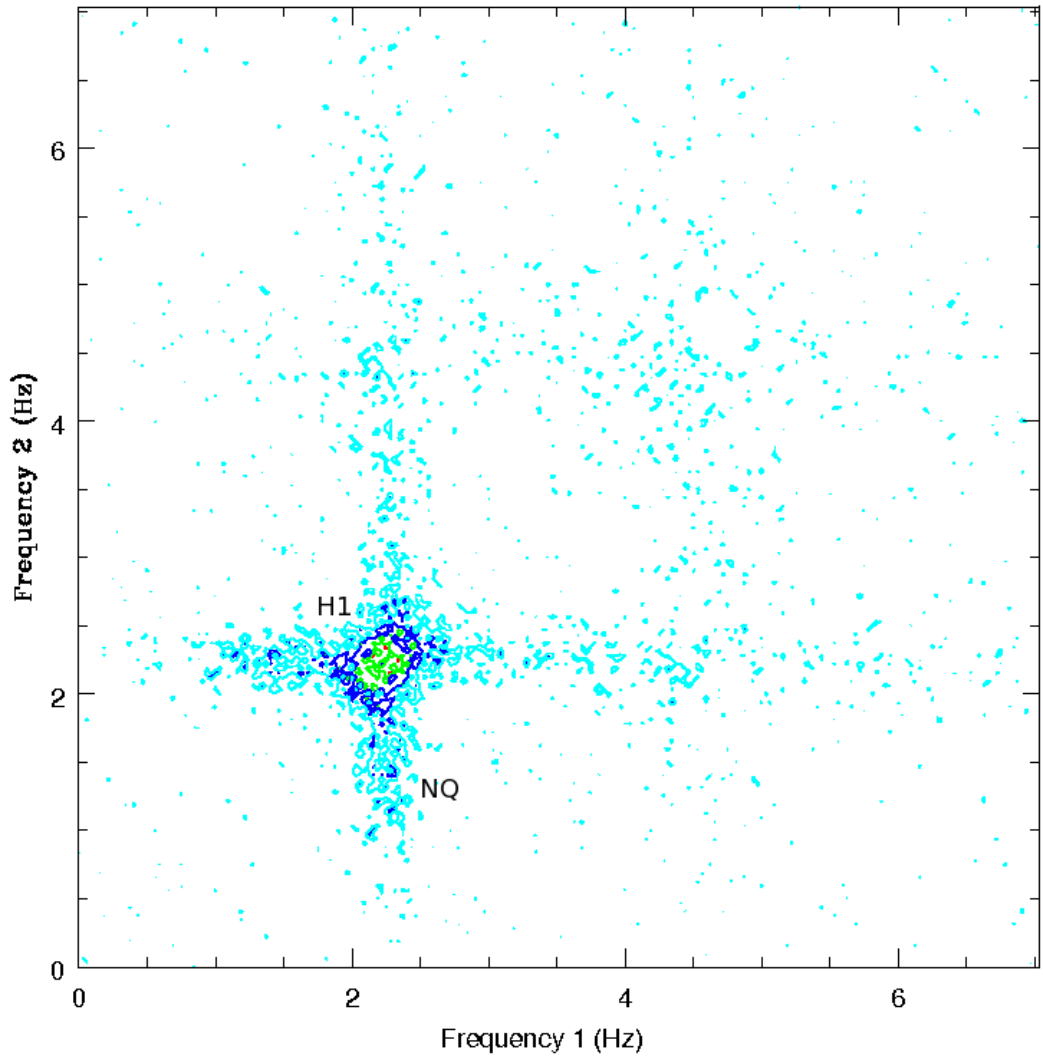}
\caption{The bicoherence plot for observation 20402-01-15-00.  The
  peak closest to H1 is the interaction between the fundamental and
  the second harmonic.  The region just to the left of the NQ label is
  the region where the noise frequency plus the QPO frequency add up
  to the second noise frequency.}
\label{bico2}
\end{figure*}

First let us examine the properties of the bispectrum for the
frequencies in which the $f_1\approx{f_2}$ is the QPO frequency and
$f_1+f_2$ is the frequency of the second harmonic of the QPO.  We take
the range of frequency bins from 69 to 76 in units of the frequency
resolution -- the region labelled H1 in figure \ref{bico2}.  The
biphases here range from $-0.16\pi$ to $-0.35\pi$, with a mean value
of $-0.29\pi$, a standard deviation of 0.05$\pi$, and a standard
deviation of the mean of 0.01$\pi$.  As the biphase is $-0.29\pi$, the
real component of the bispectrum is positive, indicating a flux
distribution skewed to the bright end of the mean.  The imaginary
component is negative, indicating a fast rise, slow decay shape to the
oscillation.

We can then look at the interactions between the QPO and the noise
component, which are strong for the triplets of frequencies of $f_1,
f_{QPO}$ and $f_1+f_{QPO}$, where $f_1<f_{QPO}$ -- the region labelled
NQ in figure \ref{bico2}.  In this case, the measurement errors are
large on the biphases, and the values span nearly the full range of
$2\pi$, so phase wrapping prevents us from using the mean and
dispersion of the biphases values directly as has been done for the
previous cases.  In order to find a mean biphase, we average the sines
of the biphases and the cosines of the biphases.  We find that the
final biphase has a value of $-0.21\pi$, and we take the variance in
the mean value of the sines and cosines of the biphases and use
standard error propagation to find a standard deviation of $0.09\pi$
and a standard deviation of the mean of $0.01\pi$.  The shape of the
light curve on these timescales is thus fairly similar to the shape of
the QPO.  As in the previous observation, while some power is present
in the bicoherence above the QPO frequency, this power is weak, so we
do not investigate the biphase there.

\subsection{30184-01-01-000: hypotenuse pattern, high freq: a nearly sawtooth pattern for the fundamental plus second harmonic}

\begin{figure*}
\includegraphics[width=6 in, angle=0]{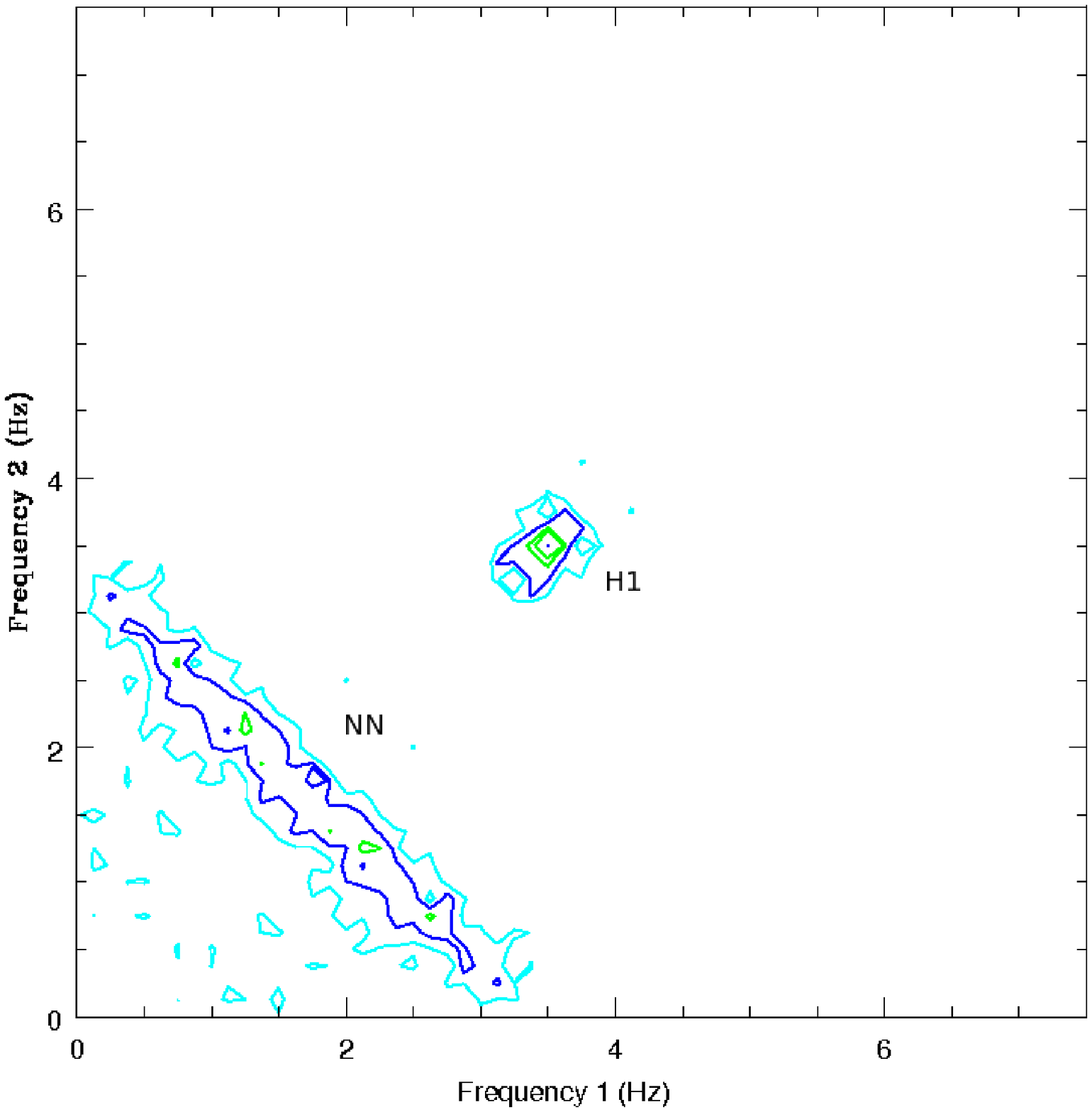}
\caption{The bicoherence plot for observation 30184-01-01-000.  The
  peak closest to H1 is the interaction between the fundamental and
  the second harmonic.  The noise region examined is the diagonal near
  the label NN -- as discussed in the text, because the QPO is
  extremely strong in this observations, the frequencies at the
  ``ends'' of the diagonals -- i.e. those with one component
  relatively near 0 frequency and one relatively near the QPO
  frequency -- are not included in the biphase computation.}
\label{bico3}
\end{figure*}

In this observation, the bicoherence shows strong power on the
timescale of the harmonic, and for the case where $f_1+f_2=f_{QPO}$,
but not for any frequency higher than the fundamental frequency of the
QPO, except at the harmonics of the QPO.  For this observation, the
QPO is at a frequency of approximately 3.4 Hz, peaking in bin 57 with
a frequency resolution of 1/16 Hz.  The bicoherence plot is given in
figure \ref{bico3}.

First, we examine the coupling between the fundamental and the
harmonic, the region marked H1 in figure \ref{bico3}.  Taking all
frequency bins from 55-59, we find that the biphasess all lie between
$1.31\pi$ and $1.55\pi$, with a mean of 1.42 $\pi$, a standard
deviation of $0.07\pi$, and a standard deviation of the mean of
.01$\pi$.  The flux distribution of the source here is thus nearly
symmetric, while the time series itself has some asymmetry.  The
biphase for the interaction between the fundamental and the second
harmonic indicates that the QPO rises quickly and decays slowly, as a
decaying sawtooth wave.  We note that the values of the squared
bicoherence in this observation, even at the harmonic, are less than
0.05, indicating that the relative phases of the fundamental and the
harmonic show substantial variation, and hence so does the shape of
the oscillation -- at the same time, the statistical significance of
the difference between the bicoherence and zero is quite strong (M11),
so the mean shape of the oscillation must be something like an inverse
sawtooth wave.

The interaction of the QPO with the noise component shows a different
behavior.  Because the QPO is extremely strong in this observation, we
must move far off the QPO centroid in order not to have the biphases
estimates affected substantially by the wings of the QPO.  We take the
means of the sines and cosines of the biphases for all cases with
$f_1,f_2<45$ in units of the frequency resolution, and $f_{QPO}$ from
55 to 59 in the same units -- this is the region marked NN is figure
\ref{bico3}.  We find that the mean biphase is $0.32\pi$ radians, with
a standard deviation of $0.2\pi$ and a standard deviation of the mean
of $0.02\pi$.  The flux distribution is thus skewed to positive
values.  The time symmetry is that of a sawtooth wave with slow rise
and fast decay -- the QPO amplitude is rising slowly and turning off
more quickly.  Thus, when we compare with obseravation 10408-01-25-00,
we see that in both observations, the interactions with the noise
component are fairly similar, while the shapes of the oscillations
themsleves are quite different.

\section{Discussion}

Since our previous paper, we have become aware of some mechanisms for
producing some of the observed bicoherence patterns in a tidy manner.
In particular, the ``hypotenuse'' pattern is reproduced very well by a
bilinear oscillator (Rivola \& White 1998; White 2009).  The bilinear
oscillator is a system described by a differential equation quite
similar to that for a simple harmonic oscillator, except that the
restoring force has a different normalization for positive and
negative displacements.  The bilinear oscillator is of interest to
engineers because it provides a good mathematical description to a
cracked or fatigued beam within a machine, and hence measuring the
bispectrum of the machine in response to being driven by vibrations
can allow the crack to be detected without taking apart the machine or
waiting for the machine to suffer a catastrophic breakdown.

While obviously cracked beams do not exist in astrophysical
situations, other types of force law with similar mathematical
dependences may exist in accretion disks.  If we find that the power
spectrum and the bispectrum of the bilinear oscillator can give a good
mathematical description of the time series we observe, then we can
focus theoretical efforts on producing physical models that are
mathematically similar to the bilinear oscillator.

Following White (2009), we calculate a bilinear oscillator which follows the
equation:
\begin{equation}
\frac{d^2y}{dt^2}+c\frac{dy}{dt}+\kappa(y)y=x(t)
\end{equation}

For a first run, we set $\kappa$ equal to 200000 for $y<0$ and to 1000000 for
$y>0$.  We set $c$ equal to 50, and we integrate over discretized time steps
of .0003 time units, with an initial value of $y$ of 200.  We drive the
oscillator with a white noise process.  The driving force has a mean
expectation value of 0.0 and an expected standard deviation of
$5.52\times10^6$.

This run produces a bicoherence plot that looks quite similar to the
hypotenuse pattern seen in the real data for observation
30184-01-01-000.  The biphase for the QPO's interaction with the
harmonic here is very nearly 0.  That is, the flux distribution is
skewed to positive values, but the time series is symmetric in time.
The same is true for the interactions of two noise components that add
up to the QPO frequency.  We next perform a simulation which is
identical to the first, except that we exchange the values of $\kappa$
with respect to $y=0$ -- i.e. we set $\kappa$ equal to 200000 for
$y>0$ and to 1000000 for $y<0$.  While the details of the time series
are different the key statistical properties measured by the biphase
are the same.  We conduct a few experiments with stronger damping.
Making the system more strongly damped decreases the bicoherence and
broadens the QPO, but does not change the biphase substantially.

It thus appears that models mathematically similar to the bilinear
oscillator are unlikely to produce the observed bispectra of X-ray
binary light curves.  There remains a catch, of course -- the
inherently oscillating parameter in a quasi-periodic oscillation is
almost certainly not just the count rate.  Models for the few Hz QPOs
we discuss in this paper include models in which the disk undergoes a
global oscillation due to the Lense-Thirring precession (e.g. Stella
\& Vietri 1998; Fragile et al. 2007), and thermal-viscous oscillations
(Abramowicz et al. 1989; Chen \& Taam 1994).  In the Lense-Thirring
pression model, the parameter oscillating is the inclination angle of
the inner accretion disk, and the X-ray count rate will be a function
of that inclination angle and may be modulated additionally, for
example, by fluctations in the accretion rate changing the surface
brightness of the disk in ways independent (or, at least, not directly
tied to) of the observer's inclination angle (some initial exploration
of this possibility has been made in Ingram \& Done 2011).  Numerical
calculations made to date do not cover enough cycles of the
oscillation to allow calculations of the biphase of oscillations from
the Lense-Thirring precession, but it is intriguing that the one
published numerical calculation which includes ray tracing does seem
to show qualitatively similar phenomenology to that see in observation
30184-01-01-000 (Dexter \& Fragile 2011).  It would also be
straightforward for the Lense-Thirring model to produce periodic
occultations of part of the accretion disk, and hence to produce a
biphase of approximately $\pi$ -- this has been considered in Ingram
\& Done (2012), although the bispectrum of the simulation has not yet
been computed.  In principle, the same type of ray tracing
calculations used in Dexter \& Fragile (2011) could be run over a
wider range of parameter space to detemine whether the biphase
properties could be matched in a manner that is also consistent with
the additional information given from other system parameters such as
the mass accretion rate.  In some cases, also, higher order
corrugation modes may be present (see e.g. Tsang \& Lai 2009).

The situation for the thermal viscous oscillation is perhaps even more
difficult to reconcile with the data.  The simulated light curves of
Chen \& Taam (1994) show slow rises, followed by quick decays of the
flux.  This would be expected to produce a biphase near $\pi/2$,
giving opposite behavior to that seen, for example, observation
30184-01-01-000.  Other models exist for explaining the low frequency
quasi-periodic oscillations in X-ray binaries, and have the
accompanied by noise components, but at present, simulated light
curves for them have not been presented which could allow us to
determine under what conditions the observed biphases might be
reproduced (e.g. Varni\'ere \& Tagger 2002; Machida \& Matsumoto 2008).

In future work, the biphase analysis may also be extended to help
develop a better understanding of the light curves of sources without
strong quasi-periodic oscillations.  In particular, the recent
development of an analytic method for calculating simulated light
curves from propagation models (Ingram \& van der Klis 2013) should
allow one to study these models efficiently to determine how well they
match the observed data in biphase.  Given that the flux distributions
of X-ray binaries are widely found to be log-normal when they are
dominated by noise components (Uttley et al. 2005), we can reasonably
expect that in all cases, the real component of the biphase will be
positive.  We can also expect that the imaginary components will be
positive as well, given that the fastest variability is expected at
the highest count rates due to the higher rate of energy generation in
the inner part of the accretion flow.  The exact value of the biphase
is likely to trace the emissivity of the accretion flow.

\section{Summary}

We have presented an introduction to the use of the biphase aimed at
astronomers wishing to apply it to time series analysis.  First, we
briefly summarize the meaning of the biphase, so that a quick look at
its value can be used to develop a quick intuition about the
properties of a timeseries.

\begin{enumerate}

\item Time series which are symmetric in time have purely real
  bispectra, and time series which have flux distributions symmetric
  about the mean have purely imaginary bispectra.

\item When the real component is positive, the flux distribution is skewed to
  positive values.  

\item Thus spiky time series like pulsar light curves will have
  biphases near 0.

\item Time series like eclipsing binary light curves will
  have biphases near $\pi$.

\item When the imaginary component is positive, the time series rises slowly
  and fades quickly.  This yields biphases near $\pi/2$.

\item When the imaginary component is negative, the time series rises quickly and fades slowly.  This yield biphases near $-\pi/2$.  

\end{enumerate}

We have also applied the biphase to several obseravtions of
GRS~1915+105.  We have found that the profile of the QPO matches well
between light curves produced by folding on the QPO period and the
predictions from the biphase data.  We have also found that the actual
values of the biphases vary widely from observation to observation.
No simple models that we have considered can reproduce the observed
biphase data.

\section{Acknowledgments}
I am grateful to Michiel van der Klis, Chris Fragile, Patricia
Ar\'evalo, Simon Vaughan, and to Paul White of the University of
Southampton's Institute for Sound and Vibration Research for extremely
interesting and valuable discussions.  I also thank the Astrophysis
Insitute of the Canary Islands for hospitality while a portion of this
work was completed.  Finally, I thank the referee, Adam Ingram, for a
report which was both prompt and helpful, and which has led to
improvements in the clarity and content of the paper.

\label{lastpage}


\begin{thebibliography}{99}
\bibitem{}Aigrain S., Favata F., Gilmore G., 2004, A\&A, 414, 1139
\bibitem{}Abramenko V.I., 2005, ApJ, 629, 1141
\bibitem{}Abramowicz M.A., Szuszkiewicz E., Wallinder F., 1989, in Theory of
  Accretion Disks, eds. F. Meyer, W.J. Duschl, J. Frank, E.Meyer-Hofmeister
  (Dordrecht: Kluwer), 141 
\bibitem{}Davies R.B., Harte D.S., 1987, Biometrika, 74, 95
\bibitem{}Dexter J., Fragile P.C., 2011, ApJ, 730, 36
\bibitem{}Edelson R.A., Krolik J.H., 1988, ApJ, 333, 646
\bibitem{}Elgar S., Guza R.T., 1985, J. Fluid Mech., 161, 425
\bibitem{}Fackrell J., 1997, PhD Thesis, University of Edinburgh
\bibitem{}Feroci M., et al., 2012, Experimental Astronomy, 34, 415
\bibitem{}Gajraj R.J., Doi M., Matzaridis H., Kenny G.N., 1998, British Journal of Anaesthesia, 80, 46
\bibitem{}Gaskell C.M., 2004, ApJ, 612, L21
\bibitem{}Hasselman K.W., Munk W. \& MacDonald G., 1963, Time Series Analysis, John Wiley: New York, M. Rosenblatt editor, p.125
\bibitem{}Haug E.G., Taleb N.N., 2011, Journal of Economic Behavor and Organization, 77
\bibitem{}Hesse K.H., Wielebinski R., 1974, A\&A, 31, 409
\bibitem{}Ingram A., Done C., 2011, MNRAS, 415, 2323
\bibitem{}Ingram A., Done C., 2012, MNRAS, 419, 2369
\bibitem{}Jiang C., et al., 2011, ApJ, 742, 120
\bibitem{}Kamionkowski M., Smith T.L., Heavens A., 2011, PhysRevD, 83, 023007
\bibitem{}Klassen A., Aurass H., Mann G., 2001, A\&A, 370, L41
\bibitem{}Loeb A., Gaudi B.S., 2003, ApJ, 588L, 117
\bibitem{}Lyutyj V.M. \& Oknyanskij V.L., 1987, AZh, 64, 465
\bibitem{}Maccarone T.J., Coppi P.S., 2002, MNRAS, 336, 817
\bibitem{}Maccarone T.J., Coppi P.S., Poutanen J., 2000, ApJ, 537L, 107
\bibitem{}Maccarone T.J., Uttley P., van der Klis M., Wijnands R.A.D., Coppi P.S., 2011, MNRAS, 413, 1819 (M11)
\bibitem{}Machida M., Matsumoto R., 2008, PASJ, 60, 613
\bibitem{}Makuch R.W., Freeman D.H., Johnson M.F., 1979, Journal of Chronic
    Disease, 32, 245
\bibitem{}Masada A., Kuo Y.-Y., 1981, Deep Sea Research, 28A, 213
\bibitem{}McComas C.H., Briscoe M.G., 1980, Journal of Fluid Mechanics, 97, 205
\bibitem{}Nowak M.A., Vaughan B.A., Wilms J., Dove J.B., Begelman M.C., 1999, ApJ, 510, 874
\bibitem{}Priedhorsky W., Garmire G.P., Rothschild R., Boldt E., Serlemitsos P., Holt S., 1979, ApJ, 233, 350
\bibitem{}Rivola A., White P., 1998, Journal of Sound and Vibration Research,
  216, 889
\bibitem{}Scaringi S., K\"ording, E., Uttley, P., Knigge, C., Groot, P. J., Still, M., 2012, MNRAS, 421, 2854
\bibitem{}Shaposhnikov N., 2012, astro-ph/1205.0748
\bibitem{}Terrell N.J., 1972, ApJ, 174L, 35
\bibitem{}Timmer J., Koenig M., 1995, A\&A, 300, 707
\bibitem{}Uttley P., McHardy I.M., 2001, MNRAS, 323L, 26
\bibitem{}Uttley P., McHardy I.M., Vaughan S., 2005, MNRAS, 359, 345
\bibitem{}van Milligen B.P., Sanchez E., Estrada T., Hidalgo C., Branas B., Carreras B., Garcia L., 1995, Physics of Plasmas, 2, 3017
\bibitem{}Varni\`ere P., Tagger M., 2002, A\&A, 394, 329
\bibitem{}Way M.J., Scargle J.D., Ali K.M., Srivastava A.N., 2012, ``Advances in Machine Learning and Data Mining for Astronomy'', CRC Press: Boca Raton
\bibitem{}White P.R., 2009, in {\it Encyclopedia of Structural Health
  Monitoring}, editors C.Boller, F.-K. Chang and Y. Fujino, Wiley:Hoboken
\bibitem{}Zweibel E. G., Yamada M., 2009, ARA\&A, 47, 291
\end{thebibliography}
\end{document}